\documentclass[aps,floats,superscriptaddress]{revtex4}
\usepackage{amsmath}
\usepackage{graphicx}
\allowdisplaybreaks[1] 

\def\GroupeEquations#1{\begin{subequations}  #1  \end{subequations}}
\def\moy#1{\left\langle #1 \right\rangle}

\def\Im{\hbox{Im}}
\def\Tr{\text{Tr}}

\def\sgn{\text{sgn}\,}
\def\parent#1{\left(#1\right)}

\def\figx#1#2{\includegraphics[width=#1]{#2}}

\def\parent#1{\left(#1\right)}
\newcommand{\empile}[2]{\genfrac{}{}{0pt}{}{#1}{#2}}

\def\matrice#1{{\begin{pmatrix}#1\end{pmatrix}}}

\def\SigmaSkel{{ \Sigma \kern -5.5pt \raise 1pt \hbox{/}}}

\def\RBZ{\text{R.B.Z}}
\def\sigmacluster{\Sigma^{\text{C}}}
\def\sigmalatt{\Sigma^{\text{latt}}}
\def\sizecluster{L}

\def\DMFT{{\it DMFT}}
\def\EDMFT{{\it EDMFT}}
\def\CDMFT{{\it CDMFT}}
\def\PCDMFT{{\it PCDMFT}}
\def\DCA{{\it DCA }}

\def\iomn{i\omega_n}
\def\ka {{k}}

\begin{document}

\title{ Cluster Dynamical Mean Field Theories}

\author{G. Biroli}
\affiliation{Service de Physique Theorique, CEA Saclay, 91191 Gif-Sur-Yvette, FRANCE}

\author{O. Parcollet}
\affiliation{Service de Physique Theorique, CEA Saclay, 91191 Gif-Sur-Yvette, FRANCE}

\author{G. Kotliar}
\affiliation{Center for Materials Theory,
Department of Physics and Astronomy, Rutgers University, Piscataway, NJ
08854, USA} 

\date{  \today}

\begin{abstract}
Cluster Dynamical Mean Field Theories are analyzed in
terms of their semiclassical limit and their causality properties,
and a translation invariant formulation of
the cellular dynamical mean field theory, \PCDMFT, is presented.
The semiclassical limit of the cluster methods is analyzed by applying
them to the Falikov-Kimball model in the limit
of infinite Hubbard interaction $U$
where they map to different classical cluster 
schemes for the Ising model.
Furthermore the Cutkosky-t'Hooft-Veltman cutting equations are
generalized and derived for non translation
invariant systems using the Schwinger-Keldysh formalism.
This provides
a general setting to discuss causality properties of cluster methods.
To illustrate the method, we prove that \PCDMFT\ is causal while the nested
cluster schemes (NCS) in general  and the pair scheme in particular
are  not.
 Constraints on further extension of these schemes are discussed.
\end{abstract}
\maketitle

\section{Introduction}

Dynamical Mean Field Theory (for a review see \cite{DMFT}) 
has been very successful in providing a non perturbative approach
to strongly correlated Fermi systems. It describes both the localized and the itinerant
limit and has yielded  non perturbative insights into the finite temperature
Mott transition \cite{DMFT}. This approach has been combined with realistic electronic structure
methods such as LDA and GW and has been applied successfully to
numerous materials \cite{RealisticComputation}.
In spite of these successes, several limitations of  
the single site {\it DMFT} approach are now apparent. For example,  the
self-energy  is $k$ independent by construction, so the method cannot
describe independent variations of the quasiparticle residue, the
quasiparticle lifetime and the effective mass. 
Furthermore, the single site nature of the method precludes it from 
treating  more exotic orders  with order parameters involving several sites,
such as dimerization, staggered flux or $d$-density  wave, and $d$-wave superconductivity.

To overcome these limitations, various extensions of \DMFT\  have been 
proposed. For a disordered  system, one can set up a functional integral formulation,
with \DMFT\  as a saddle point, leading to a natural loop expansion
\cite{MCA}. 
A formulation of these ideas for a clean system is still lacking.
A different extension, is the Extended \DMFT\  \cite{EDMFT}, in which the \DMFT\  Ansatz
is applied simultaneously to bosonic and fermionic degrees of freedom.
This approach does not describe a $k$-dependent self-energy but incorporate
diagrams involving longer range interactions into the \DMFT\  equations. 

A different idea is based on truncations of the Baym Kadanoff
functional.
In the full theory, this is a functional of the full Green functions.
\DMFT\  is obtained by restricting it to local Green functions
only, setting the non-local Green functions equal to zero. 
A natural extension is to restrict the functional to local and nearest
neighbor Green functions.
This pair scheme  was introduced independently by Ingersent and
Schiller \cite{Ingersent-Schiller},
and by Georges and Kotliar\cite{DMFT}.
The latter authors observed non causal
behavior in an Iterated Perturbation Theory (IPT)  solution of the Hubbard model, but the origin of this
problem was not elucidated. Zarand et al. suggested
\cite{Zarand-Pair-Scheme} that the pair  cluster
approach was causal, and that the difficulties encountered in the solution
were related to the impurity solver rather than to the scheme itself,
but no conclusive proof of this statement was presented.
This  pair method requires a simultaneous solution of multiple site
impurity problems.
 
A different direction was pursued by Jarrell and collaborators
with the introduction of Dynamical Cluster Approximation
(\DCA) \cite{DCA-all}, whose main idea is to discretize uniformly momentum space.
This approach was shown to be manifestly causal.
It can also be  formulated in real space  \cite{Kotliar-Biroli} (See
also Appendix {\ref{AppendixDCA}).
A different approach, the cellular \DMFT\  or \CDMFT,
motivated by applications to electronic  structure, was introduced
in \cite{CDMFT}. In this approach, the many body problem is truncated by
introducing a finite basis set  of orbitals to truncate the self energy.
It introduces the cluster self energy and the  lattice self energy as independent entities.
This method was tested in a soluble model
\cite{Kotliar-Biroli}, in the 
Falikov-Kimball model \cite{Jarrell-Falikov-Kimball} 
and in the one dimensional Hubbard model \cite{Kotliar-Venky-1d}.
These papers developed the \CDMFT\  ideas from a real space perspective.
In this paper we develop this  approach  from a momentum space perspective,
to maintain periodicity in the self consistency   
equation. The importance of including this periodicity was underlined by
Lichtenstein \cite{Lichtenstein-private,Lichtenstein-Katsnelson}
arguing that the lack of periodicity of \CDMFT\  could unfavor phases, like the d-wave
superconductivity, in which the order parameter lives on links.
Even though this point has not yet been elucidated, it is certainly desirable to have a generalization of CMDFT that is
translation invariant and causal  in its use of the lattice self-energy in the self consistency condition.

Cluster \DMFT\  methods have not yet reached the level of
understanding of their single site counterparts. 
While the single site \DMFT\   can be unambiguously formulated,
cluster \DMFT\  methods are more diverse and therefore require   more detailed methodological
investigation, since the virtues and limitations of the various cluster
schemes are not apparent yet.  
This paper is a contribution in this direction.

There are several important principles that a cluster method
should satisfy:
{\it i)} Given that cluster \DMFT\  approximations are intrinsically basis
dependent, a cluster method
should be formulated in a general basis set. This flexibility is important,
because for a given problem, one could carry out the cluster \DMFT\  study in the basis
which is most suitable for the system in question.
{\it ii)} It should have an effective action formulation, namely it should target
the calculation of a specific correlator function.
{\it iii)} It should yield causal Green functions.
{\it iv)} It should be able to capture the various order, including
those which break the translation invariance. 
{\it v)} It should converge rapidly as  a function of the discretization
parameter for the observable  that one is targeting.
It is possible, that different cluster schemes may converge
better for different observables.
A better understanding of these elements of a cluster method is desirable.

In section \ref{sec.desc.methods},  we present our translation invariant
generalization of \CDMFT\  where
the lattice self energy participates in the self consistent equation.
This is a new cluster scheme, and we formulate it in a way which
allows a comparison with \CDMFT\   and \DCA. 
We also discuss in section \ref{sec.nested.clusters} a different class of schemes, the {\it nested cluster schemes},
which require the simultaneous consistent solutions of impurity models
of different size.

In section \ref{section.classicallimit}, in an attempt to clarify the nature of the various cluster approximations,
we study their classical limit in the Ising limit of the
Falikov-Kimball model. The various schemes then reduce to classical cluster approximations
to the Ising model. This analysis elucidates their
physical content at the classical level and it allows a simple and clear comparison.
This study complements  other  comparisons of cluster schemes against
more exact treatments for specific models \cite{Kotliar-Biroli,Jarrell-Falikov-Kimball}. 
The classical limit of \EDMFT\  was discussed in \cite{Pankov-EDMFT-Classical}.

In section \ref{section.causality}, we analyze the issue of causality of cluster methods
from a diagrammatic perspective using the Cutkovsky-t'Hooft-Veltman
rules. We rederive them for non translation invariant cases using the
Schwinger-Keldysh formalism. We end up with a general
setting to analyze the causality of cluster schemes. Within this
framework we show that the pair scheme is not causal, and elucidate
the origin  of the problem. Furthermore
we provide a simple proof of the causality of \CDMFT\  and \DCA\  (which had
already been proved by other methods) and  justify the causality
of the periodic generalization of \CDMFT.
Finally we make a relationship with the early paper of Mills et al.
on disordered systems \cite{Mills}: in fact the coherent potential approximation (CPA) is a particular case
of a \DMFT. The origin of violations
of causality, and their possible cure for the generalizations of CPA were clarified by Mills \cite{Mills}. 

\section{Description of the cluster methods}\label{sec.desc.methods}

A fairly general   model of strongly correlated
electrons contains hopping and interaction
terms. It is  defined 
on a lattice of ${\cal  N}$ sites in $d$ dimensions, and we divide the 
lattice in $({\cal  N}/L)^{d}$ cubic clusters of $S_{c}= L^{d}$ sites (more general
forms can also be considered). We denote with 
${\mathbf{e}}_{i}$ the internal cluster position and with ${\mathbf{R}}_{n}$ 
the cluster position in the lattice (therefore the position
of the $i$-th site of the $n$-th cluster is $R_{n}+{r}_{i}$).
The lattice Hamiltonian  is expressed in terms of 
fermionic operators $c^{\dagger }_{R_{n},\alpha }$ and $c_{R_{m},\beta }$
and can be written  as:
\begin{equation}\label{hamiltonianBasic}
H=\sum_{n,\alpha ,m,\beta }t_{\alpha ,\beta } (R_{n}-R_{m})c^{\dagger }_{R_{n},\alpha }c_{R_{m},\beta }
+\sum_{n,\alpha ,m,\beta ,n',\alpha ',m',\beta '}
U_{\alpha ,\beta ,\alpha ',\beta '}(\{ {R}\})
c^{\dagger}_{{R}_{n},\alpha }c_{{R}_{m},\beta }c^{\dagger }_{{R}_{n'},\alpha'}c_{{R}_{m'},\beta '} 
\end{equation}
where
$\alpha =i,\sigma $ and $\sigma $ is an internal degree
of freedom (i.e. a  spin,   spin-orbital or band index).
Note that  in all this paper, we will denote the position and momentum on the (original)
lattice with lowercase letters ($r$ and $k$) and the position and
momentum of the superlattice with uppercase letters ($R$ and $K$).
Different dynamical cluster methods for strongly correlated have been
introduced in order to obtain an approximate solution of
(\ref{hamiltonianBasic}) able to capture the effect of short range
(dynamical) correlation and to describe the self-energy k-dependence.

In the following, we introduce a new scheme, the Periodized Cluster
Dynamical Field Theory (\PCDMFT), formulate \CDMFT\  and \DCA\  in real
and $k$ space and the {\sl nested cluster schemes (NCS)}, {\it i.e.} the pair
scheme and its generalisations. 
\CDMFT\  \cite{CDMFT} is a real-space cluster : the lattice is divided in a
superlattice of cells and the scheme is basically the \DMFT\  equations
on the superlattice \cite{DMFT}.
On the other hand, \DCA\cite{DCA-all}  is a reciprocal space cluster, 
where the self-energy in momentum space is approximated by step
function around a few points, which we will denote by $K_{c}$ and are
identified to the momentum of the (periodic) cluster.
We will show that \PCDMFT\  is a natural generalization of both schemes,
from a real space and a $k$-space perspective.

\subsection{Real space perspective}
Cluster methods for strongly correlated
electrons (\CDMFT, \DCA, \PCDMFT) can be divided in two steps. 
(Nested cluster schemes will be considered separately).

The first step is the computation of  the local cluster propagator
$G_{c\sigma \mu \nu }(\tau)$ and the cluster self-energy
$\sigmacluster$  from an effective action 
containing a Weiss dynamical field $G_{0,\sigma  \mu \nu }^{-1} (\tau ,\tau')$
and the intra-cluster interaction :
\GroupeEquations{\label{Eq.Cluster.2}
  \begin{align}
S_{\text{eff}} &= - \iint_{0}^{\beta } d \tau  d \tau'
c^{\dagger }_{\sigma \mu  } (\tau )
G_{0,\sigma  \mu \nu }^{-1} (\tau ,\tau')
c^{ }_{\sigma \nu  } (\tau' )
+ \int_{0}^{\beta } d \tau
U_{\alpha \beta \gamma \delta }(R=0)
(c^{\dagger }_{\alpha } c_{\beta }
c^{\dagger }_{\gamma  } c_{\delta  })(\tau )
\\
G_{c\sigma \mu \nu }(\tau) &= -\moy{Tc_{\sigma \mu }(\tau)
 c_{\sigma \nu }^{\dagger }(0)}_{S_{\text{eff}}}\\
\label{ClusterEqGal.DefSelf}
 \sigmacluster &=  G_{0}^{-1} - G_{c}^{-1}
 \end{align}}
where $\mu,\nu$ are indices for cluster sites. In the following, we
will often concentrate on square clusters of linear size $L$ on a $d$-dimensional square
lattice, although many of the results can be generalised easily.
Hence,  $\mu ,\nu$ will also denote the position of the cluster sites on
the lattice : $\mu ,\nu  \in {\cal  C} =\{0,\dotsc ,L-1 \}^{d}$ where
$d$ is the dimension (intersite distance is
normalised to one).
We will denote by $S_{c}$ the cluster's size and by ${\cal  C}$ the
set of cluster points;  $\sigma$ is the spin index as
mentioned earlier and $\alpha,\beta,\gamma,\delta$
are double indices gathering the spin and the cluster index.

The second step consists in recomputing the Weiss field using the value
of the self-energy obtained by the first step and then iterating until
convergence is reached. The real difference between cluster schemes
is how the second step is performed.

\CDMFT\ is a direct generalization of \DMFT\  to cluster {\it
in real space} and consists simply in rewriting the \DMFT\  equations in a matrix
form ($\mu ,\nu$ indices are omitted) taking as elementary degrees of freedom all the cluster fermionic
degrees of freedom:
  \begin{align}
\label{CDMFT}
 G_{0}^{-1} (i\omega_n ) &= \left(\sum_{K\in R.B.Z.} \frac{1}{i\omega_n + \mu
 - \hat{t} (K) - \sigmacluster (i\omega_n ) } \right)^{-1}  +
\sigmacluster (i\omega_n )
    \end{align}
where $\hat t_{\mu \nu } (K)$ is the hopping expressed in the superlattice
notations, with $K$ in the Reduced Brillouin Zone (R.B.Z.) of the superlattice.
(See Equation \ref{App.DCA.t.supernotations} in Appendix). Note that from now on the
sum over $K$ means always the normalized sum.
When convergence is reached and the cluster self-energy has been obtained,
the translation invariant lattice self-energy 
$\sigmalatt$ ($(i,j)$ denotes a site on the original lattice) is computed by the
formula:
\begin{equation}\label{def.Sigmalatt.1}
\sigmalatt_{i-j} (\omega ) = \frac{1}{S_{c}}
\sum_{
\empile{\mu ,\nu  \in {\cal  C }} {{\mu} - {\nu}  = i-j}} 
\sigmacluster_{\mu \nu } (\omega )
\end{equation}
Other estimators (that still preserve causality) have been introduced
in \cite{Kotliar-Biroli}. Note that
in \CDMFT, the cluster  and the lattice self-energy are two different
quantities \cite{CDMFT} : one has to first solve the cluster problem,
and only at the end to compute the lattice self-energy $\sigmalatt$. 
{\sl In \CDMFT, $\sigmalatt$ does not enter the self-consistency
condition}.

\DCA\  is more naturally formulated in Fourier space \cite{DCA-all}.
However it can also be recovered from a real space perspective
\cite{Kotliar-Biroli} 
 just changing the hopping matrix in eq. (\ref{CDMFT}) to  (See also
Appendix \ref{AppendixDCA}) :
\[
\hat t^{DCA}_{\mu \nu} (K) \equiv  t_{\mu \nu  } (K)\exp (-iK \cdot ({\mu  }-{\nu  }))
\]

In this case if the
self-energy is a cyclic matrix (translation invariant within the
cluster) then eq. (\ref{CDMFT}) reduces to (upon diagonalisation) :
  \begin{align}
\label{def.DCA}
 G_{0}^{-1} (K_{c},i\omega_n) &= \left(\sum_{K\in R.B.Z.} \frac{1}{i\omega_n + \mu
 - {t} (K+K_{c}) - \sigmacluster (K_{c},i\omega_n ) } \right)^{-1}  +
\sigmacluster (K_{c},i\omega_n )
    \end{align}
that is the standard \DCA\  equation introduced in \cite{DCA-all}. Note that
if the self-energy is cyclic then the Weiss field computed by
(\ref{def.DCA}) will be cyclic, so this property is preserved within the 
self-consistent loop.
In \DCA, contrary to \CDMFT,
there is no distinction between the cluster and the lattice
self-energy. Because it is formulated in $k$-space, \DCA\  is
also naturally translation invariant.

We are now ready to define the new scheme, \PCDMFT\  for Periodized
\CDMFT. The simplest definition is to take $\sigmalatt$ in place of
$\sigmacluster$ in the self-consistency condition. This is very close
to the scheme proposed by Lichtenstein and Katsnelson in \cite{Lichtenstein-Katsnelson}, with the big
difference that \PCDMFT\  is causal, as will be proved below.
Thus the equation relating the self-energy to the Weiss field are:
\GroupeEquations{\label{Eq.Cluster.1}
  \begin{align}
\Sigma_{latt} (k,i\omega_n )&=\frac{1}{S_{c}}\sum_{\mu ,\nu \in {\cal C}}
\sigmacluster_{\mu \nu } (i\omega_{n}) \exp (-ik \cdot ({\mu }-{\nu }))\\
G_{\mu \nu  }&=
\sum_{k} 
\dfrac{ e^{-i k \cdot {\mu }} e^{i k \cdot {\nu }}}{i\omega_n + \mu
 - { t} (k) - \sigmalatt (k,i\omega_n ) } \\
G_{0}^{-1}&= G^{-1}+ \sigmacluster  
 \end{align}}
where $k$ is in the Brillouin zone of the original lattice.

The three schemes (\CDMFT, \DCA, \PCDMFT) can be summarized into
the same matrix equations : Eq. (\ref{Eq.Cluster.2}) and  
\begin{equation}\label{ClusterEqGal.SelfConsistency}
G_{0}^{-1} (i\omega_n ) = \left(\sum_{K\in R.B.Z.} 
\biggl ( {i\omega_n + \mu
 - {\hat t}_{S} (K) - \Sigma_S (K,i\omega_n ) } 
\biggr )^{-1}
\right)^{-1}  +
\sigmacluster (i\omega_n )
\end{equation}
where the difference between the three schemes is enclosed in
the value of $t_{S}$ and of $\Sigma_{S}$ that enter in the self-consistency condition.

Let us now turn to translation invariance breaking phases, where 
translation invariance is conserved 
on the superlattice (e.g. antiferromagnet, charge
density, ``stripes'' if the cluster is big enough).
\CDMFT\  can describe such
an order by construction since it does not require the  translation
invariance.
When solved numerically (e.g. with Quantum Monte Carlo method), the translation
invariant solutions are often found to  be instable towards the ordered one
(we have explicitely encountered this phenomenon for example for AF and charge density wave).

\DCA\  and \PCDMFT\ require translation invariance, therefore they need to
be generalized to handle such an order.
For \DCA\ there are two solutions for this problem, for example for
antiferromagnetic order :
{\it i)} keep a reciprocal space formulation with a Reduced Brillouin
Zone and introduce some correlation between $k$ and $k+Q$, with $Q=
(\pi ,\pi)$ \cite{MaierPhD}; {\it ii)} use the {\it real space formulation}
introduced in \cite{Kotliar-Biroli}, where translation invariance in
the cluster {\it can} be broken, and look for an antiferromagnetic solution. 
It is shown in Appendix \ref{AppendixDCA} that the two approaches are
equivalent.
However, {\it i)} {\it requires to  anticipate}  the appearance of ordered phase, 
i.e. to adapt the cluster scheme for the order to be described :
one needs to use a special setup for antiferromagnetic order, another
for a more complicated order, whereas {\it ii)} {\it does not require
to  anticipate the order }: it will show up automatically solving 
the real space \DCA\  equations with no need to generalize the
scheme (provided that the cluster is big enough to contain the unit
cell).  The same numerical code will produce a translation
invariant solution, or an antiferromagnetic one, a stripe-like one.
In particular, since translation
invariant solution are often found to often be numerically instable,
one can not miss a ordered phase with this approach.
Therefore, the {\it real space formulation} of DCA is the best
solution from a practical point of view.

The \PCDMFT\  is unfortunately more complicated to generalize.
In the following, we will focus on square clusters on the square
lattice.
It is useful to introduce a slightly more general formula 
for $\sigmalatt$ for a bipartite lattice. The lattice self-energy is a sum of the cluster
self-energy put at all possible positions on the lattice.
We can rewrite  (\ref{def.Sigmalatt.1}) in a more transparent way, as
a sum over all possible shifts of the cluster   :
\begin{equation}\label{Def.SigmaLatt}
\sigmalatt_{ \sigma, \mu ;\sigma', \nu } (K)
= \frac{1}{S_{c}}
\sum_{ { \delta }\in \{0,\dotsc ,L-1 \}^{d}}
e^{-i K \cdot \parent{
\lfloor \frac{{\mu} + {\delta}}{\sizecluster} \rfloor
-
\lfloor \frac{{\nu} + {\delta}}{\sizecluster} \rfloor
}L}
\sigmacluster_{ \sigma , \overline{\mu +{\delta}};\sigma', \overline{\nu +{\delta}}}
\end{equation}
where $\lfloor x \rfloor$ is the integer defined by $\lfloor x \rfloor
\leq x < \lfloor x \rfloor +1$ (for a vector it has to be understood
component by component), $d$ is the dimension,
${\delta}$ is a $d$-dimensional shift vector,
the bar denotes the modulo reduction by $L$
component by component (adding $\delta$ is a circular shift in the cluster).

The idea of the generalization is then simple. In a phase that breaks
translation invariance (we take antiferromagnet as an example), we have
multiple solutions (two for AF), denoted by an index $\alpha$ ($\alpha
=1,2$ for AF). In order to respect the order, we
need to compensate the shift in the cluster by the change of solution
: in the AF example, use one solution on one sublattice, another on
the other sublattice.
The formula is then 
\GroupeEquations{\label{Def.SigmaLatt.AF}
\begin{align}
\sigmalatt_{ \sigma, \mu ;\sigma', \nu } (K)
&= \frac{1}{S_{c}}
\sum_{ { \delta }\in \{0,\dotsc ,L-1 \}^{d}}
e^{-i K \cdot \parent{
\lfloor \frac{{\mu} + {\delta}}{\sizecluster} \rfloor
-
\lfloor \frac{{\nu} + {\delta}}{\sizecluster} \rfloor
}L}
{\sigmacluster}^{\alpha (\delta )}_{ \sigma , \overline{\mu +{\delta}};\sigma', \overline{\nu +{\delta}}}
\\
\alpha ({\delta }) &=
\begin{cases}
1   &  \text{ if } \sum_{i=0}^{d} \delta_{i} = 0 \ [2] \\
2  & \text{ if } \sum_{i=0}^{d} \delta_{i} = 1 \ [2] 
\end{cases}
\end{align}
}
For the AF phase and when the original model is $SU (2)$ invariant,
the solution $\alpha =2$ is simply obtain from $\alpha =1$ by a spin-flip. This does not
need to be true in general, either for more complicated order or 
the Falikov-Kimball model that we will use in Section
\ref{section.classicallimit}.
Let us note however that this generalization is more complex than for
the \DCA\ scheme from a practical perspective : the expression of $\sigmalatt$ strongly depends on
the order to be described and on the form of the cluster.

\subsection{Reciprocal space perspective. $\Phi$ derivation.}

We now explore the relation between \PCDMFT, \CDMFT\  and \DCA\  from the
reciprocal space perspective. In order to do this, we
use the generating functional formulation of \DMFT\  \cite{DMFT}. 
Let us recall that one can define a functional $\Gamma
(G)$:
\begin{equation}\label{BKfunctional}
\Gamma(G)=\Tr \log G-\Tr G_{0}^{-1}G+\Phi (G)
\end{equation}
where $G_{0}= (i\omega_{n}-t+\mu )^{-1}$ is the bare propagator and
$\Phi (G)$, the Baym-Kadanoff functional, is the sum of all
the vacuum two-particle irreducible diagrams constructed with the
propagator $G$ and the interaction vertices.
The solution of the stationarity equation of (\ref{BKfunctional})
is the real propagator of the full interacting theory. 
\DMFT, as well as its cluster generalizations, can be seen as an approximation
onto $\Phi (G)$. Indeed one obtains the \DMFT\  approximation restricting
$\Phi (G)$ only to single site propagator (equating to zero all the non
single site propagator), i.e. $\Phi_{DMFT}(G)=\sum_{i}\Phi (G_{ii}|G_{ij}=0)$
or, equivalently, neglecting the momentum conservation at the vertices.
These two procedures, that are equivalent at the single site level,
represent two different routes in order to obtain cluster
generalization of \DMFT.

\CDMFT\  can be obtained by a natural real space extension of the
\DMFT\  approximation on $\Phi $:
\begin{equation}\label{CDMFTphi}
\Phi_{CDMFT}(G)=\sum_{R}\Phi (G_{\mu , R;\nu,R}|G_{\rho,R;\lambda 
R'}=0)
\end{equation}
It's easy to show that the stationarity equation corresponding to this
choice of $\Phi $ gives back the \CDMFT\  equations. Note that 
the summation of the infinite series of diagram of $\Phi_{CDMFT}(G)$ is
performed by the cluster impurity solver like in the \DMFT\  case.\\
\DCA\  is formulated more naturally as an approximation
in Fourier space: instead of neglecting completely the momentum vertex
conservation one puts a coarse-grained delta inside the complete
diagrammatic series.
\begin{equation}\label{DCAPhi}
\Phi_{DCA}(G)=N_{ss}\left. \Phi (G (k))\right|_{U (k_{1},k_{2},k_{3},k_{4})=U_{DCA}(k_{1},k_{2},k_{3},k_{4})}
\end{equation}
where $U_{DCA}(k_{1},k_{2},k_{3},k_{4})=\delta_{K_{c}
(k_{1})+K_{c} (k_{2}),K_{c} (k_{3})+K_{c} (k_{4})}/N_{ss}$,
$N_{ss}$ is the number of clusters (number of sites divided number of
sites per cluster) and
$K_{c} (k)$ is a function that for each $k$ gives the center
of the $[\frac{\pi }{L},\frac{\pi}{L}]^{d}$ cube to which 
$k$ belongs \cite{DCA-all}.\\
\PCDMFT\  has been introduced previously from the real space point
of view like a natural generalization of \CDMFT\  in which one puts the
lattice self-energy inside the self-consistent loop (still preserving 
the causality properties of \CDMFT). In the following we shall show
that, from
the functional point of view, \PCDMFT\  can also be seen {\it at the same time } 
as a translation invariant formulation of \CDMFT\  and as a generalization of \DCA.
Up to now we restricted the discussion to the standard completely
localized basis set for simplicity. However, \CDMFT\  and \PCDMFT\  
can be formulated in a general basis set. This flexibility is important,
because for a given problem, one could carry out the analysis in the basis
which is most suitable for the system in question. As a consequence in the
following we will derive \PCDMFT\  for a general basis.

We shall show that there are two different procedures leading
to the same formulation of \PCDMFT. Let us
focus first on the one which show that \PCDMFT\  is the natural
translation invariant formulation of \CDMFT.
Call ${\varphi }_{R\alpha }$ the basis function 
used to define the cellular \DMFT\  \cite{CDMFT} and 
$\widetilde{\varphi}_{R\alpha }$ their Fourier transform. Note that they are
normalized in such a way that $\sum_{i}|{\varphi }_{R\alpha }(i)|^2
=1,\sum_{k}|\widetilde{\varphi }_{R\alpha }(k)|^2 =1$. The particular
case of the standard completely localized basis set corresponds to
$\widetilde{\varphi }_{R\alpha }(k)=\exp (-ik\cdot x_{R\alpha
})/\sqrt{N}$ ($N$ is the total number of sites).
\CDMFT\  in a general basis is obtained: {\it i)} keeping only the
intra-cluster interaction: $U_{R_{1}\alpha ,R_{2}\beta ,R_{3}\gamma
,R_{4}\delta }\rightarrow U_{R\alpha ,R\beta ,R\gamma
,R\delta }$ and {\it ii)} making the approximation (\ref{CDMFTphi})
discussed above to the Baym-Kadanoff functional.\\
\PCDMFT\  can be obtained making the approximations {\it i), ii)}  and {\it imposing} the
translational invariance of the original problem. This is performed
expressing the propagator in the new basis set in terms of the
{\it translational invariant propagator} in the original basis set:
\begin{equation}G_{R\alpha,R'\beta } (\omega ) =\sum_{k}
\widetilde{\varphi }_{R\alpha }(k)^{\ast } \hat G(k,\omega ) \widetilde{\varphi }_{R'\beta }(k)
\label{gc}
\end{equation}
Plugging this expression inside the functional leading to \CDMFT\  we get
the \PCDMFT\  functional:
\begin{equation}
\Gamma (\hat  G(k))=\Tr \log (\hat G)+\Tr 
\bigl ( (i\omega_{n}-t (k)+\mu)\hat G\bigr )
+N_{ss}\Phi_{PCDMFT} \bigl (\widetilde{U},\hat G\bigr )
\label{phi}
\end{equation}
where $\Phi $ is the Baym\ Kadanoff functional expressed in terms of
$\hat G(k)$
and obtained replacing the original matrix of interactions
$U ( k_{1},k_{2},k_{3},k_{4})$ with 
\begin{equation}\label{UPCDMFTfunctional}
\widetilde{U}( k_{1},k_{2},k_{3},k_{4})=\sum_{\alpha ,\beta, \gamma
,\delta }U_{0\alpha ,0\beta ,0\gamma
,0\delta }%
\widetilde{\varphi }_{0\alpha } (k_{1})\widetilde{\varphi }_{0\beta }^{* }(k_{2})\widetilde{%
\varphi }_{0\gamma }(k_{3})\widetilde{\varphi }_{0\delta }^{* }(k_{4})
\end{equation}
where we made use explicitly of the translation invariance to put $R=0$.

Extremizing this functional with respect to $\hat G(k)$
gives the \PCDMFT\  equations together with the form of the lattice self
energy: 
\GroupeEquations{
\begin{align}\label{PCDMFTeq1}
 G_{R\alpha ,R\beta } (\omega ) &=\sum_{k}\widetilde{\varphi }_{R\alpha }^{\ast } (k)
\frac{1}{\omega-\epsilon _{k}-\Sigma_{latt} (k,\omega )} \widetilde{\varphi }_{R\beta }(k)\\
\Sigma_{latt} (k,\omega )&=\sum_{\alpha \beta R}\widetilde{\varphi }%
_{R\alpha }^{\ast }(k)\Sigma _{\alpha \beta }^{c}\widetilde{\varphi }_{R\beta
}(k)\label{PCDMFTeq2}
\end{align}
}
where $\Sigma _{\alpha \beta }^{c}$ is the self-energy obtained from
a cluster impurity problem characterized by a propagator $G^{c}_{\alpha
,\beta }=G_{R\alpha ,R\beta }$ and the interaction matrix $\widetilde{U}_{\alpha \beta \gamma \delta }$.\\
As discussed before there is another procedure to get the functional formulation of \PCDMFT\  encoded in 
(\ref{phi}). As the \DMFT\  approximation can be obtained
just by neglecting the momentum conservation at the vertex, one can
interpret (\ref{UPCDMFTfunctional}) as improvement to the \DMFT\
approximation in which the vertex is replaced by $\tilde{U}( k_{1},k_{2},k_{3},k_{4})=\sum_{\alpha ,\beta, \gamma
,\delta }\tilde{U}_{\alpha ,\beta ,\gamma
,\delta }%
\widetilde{\varphi }_{0\alpha }(k_{1})\widetilde{\varphi }_{0\beta }(k_{2})\widetilde{%
\varphi }_{0\gamma }(k_{3})\widetilde{\varphi }_{0\delta }(k_{4})$
where $\tilde{U}_{\alpha ,\beta ,\gamma
,\delta }=U_{0\alpha ,0\beta ,0\gamma
,0\delta }$.\\
At this point a natural question is why we have chosen this particular
$\tilde{U}$ and, more interestingly, could other choices
lead to a better approximation and what is the procedure to find the
``best'' $\tilde{U}$?
Unfortunately we don't have clear answers to these questions. 
A partial answer to the first one is that the choice of $\tilde{U}$
leading to \PCDMFT\  is the one that we obtain applying the least square
minimization to $U ( k_{1},k_{2},k_{3},k_{4})-\tilde{U}(
k_{1},k_{2},k_{3},k_{4})$. So in a certain sense it provides the best
approximation to $U$ within the chosen basis set. It's however
important to notice that is far from clear that this is a good
criterion to select the ``best'' $\tilde{U}$.
With respect to these remarks is particularly interesting to note
that \DCA\  can be obtained taking $\tilde{U}_{\alpha ,\beta ,\gamma
,\delta }=U_{0\alpha ,0\alpha  ,0\alpha 
,0\alpha }\delta_{\alpha ,\beta }\delta_{\gamma ,\delta
}\delta_{\delta,\gamma }$ (in the case of a complete diagonal interaction on
the original lattice) and $\widetilde{\varphi }_{R\beta }(k)=\exp (-iK_{c}
(k)\cdot x_{R\beta })/\sqrt{N}$ \cite{DCA-all}. 

This suggests interesting interpolation between
\DCA\  and \PCDMFT. A particularly simple one consists in dividing the
Brillouin zone in $l^{d}$ squares (like in \DCA\  for a cluster of
linear size $l$) and taking $\widetilde{\varphi }_{R\mu  }(k)=\exp (-iF_{l}
(k)\cdot x_{R\mu })/\sqrt{N}$ where $F_{l} (k)$ is a function such 
that for each $k$ gives the center
of the $[\frac{\pi }{l},\frac{\pi }{l}]^{d}$ cube to which $k$ belongs
(note that $\mu =1,\dotsc ,L_{c}$). When $l=L_{c}$ one gets \DCA,
whereas when $l=\infty $ one gets \PCDMFT. 

Let us finally note that this functional derivation of \PCDMFT\  is crucial to prove simply its
causality properties. Indeed this scheme is causal because it is a 
composition of two causal steps. Clearly starting from a causal cluster self-energy
one obtains a causal cluster propagator through
(\ref{PCDMFTeq1},\ref{PCDMFTeq2}). Furthermore we will show in Section
\ref{section.causality} that plugging
the causal cluster propagator inside the diagrammatic series
corresponding to the \PCDMFT\  functional gives a causal cluster
self-energy.

\subsection{Nested cluster schemes}\label{sec.nested.clusters}

Another natural generalization of DMFT to the cluster setting is
to apply the ideas connected to the {\sl cluster variation method} (CVM)
of classical statistical mechanics to the Baym Kadanoff functional
\cite{kikuchi,morita}.
 These schemes are also natural generalizations of
the $n$ site CPA \cite{butler} to interacting electrons.
The approach is defined by selecting a set of maximal (namely
they are not included in  each other) clusters of sites.
We denote  by $\Gamma$   the set of maximal clusters together with all its
subclusters, and by $ \Phi_{\alpha}$  the {\it restriction} of the Baym Kadanoff
functional to $G_{\alpha}(i,j)$ ,
with $G_{\alpha}(i,j)=G(i,j)$ if $i$ and $j$ belong to $\alpha$ and $G_{\alpha}(i,j)=0$
otherwise.
${\tilde{ \Phi}}_{\alpha} $ is defined recursively in terms of the $\Phi_{\alpha}$
 by  $\Phi_{\alpha} = \sum_{\beta \subseteq \alpha } {\tilde \Phi}_{\beta}  $
which can be inverted by the Moebius formula
$ \tilde \Phi_{\alpha} = \sum_{\beta \subseteq \alpha } (-1)^{n_{\beta} -n_{\alpha}}{ \Phi}_{\beta}  $
with $n_{\alpha}$  the number of sites of cluster $\alpha$.
An approximation scheme is uniquely fixed once a set of maximal clusters is chosen.
If the chosen set is invariant under translations, e.g.
the set of all plaquettes, we construct a cluster scheme which
is manifestly translation invariant by truncating the full Baym Kadanoff functional
\begin{equation}
\Phi \approx \sum_{\alpha \epsilon \Gamma}{\tilde{\Phi}}_{\alpha}
\end{equation}

Differentiation of $\Phi$ yields a translational invariant self energy,
which requires the solution of several impurity problems.
The subclusters of a maximal cluster are generally related among
each other by the operations of the crystal group, and fall
into different equivalence classes. To compute the lattice self
energy  one needs to
solve several impurity models.  One for each
representative of inequivalent subclusters   of the maximal cluster.

When we take as a set of maximal subclusters
the set of all nearest neighbor pairs of the lattice we
obtain  the pair scheme (or two impurity scheme) \cite{DMFT,Ingersent-Schiller} which we
discuss in more detail for completeness.
It is convenient to go back to a lattice notation
where lattice sites are denoted by $i$ and $j$ and not by $R \mu$.
The approximation to the $\Phi$ functional can be written in terms of $\Phi_{1}$ and $\Phi_{2}$ which are
the Baym Kadanoff functionals of a one and a two impurity problem.
\begin{equation}
\Phi_{pair} =(1-z)\sum_{i}\Phi _{1}[G_{ii}]\,+\,\sum_{<ij>}\Phi
_{2}[G_{ii},G_{jj},G_{ij}]
\end{equation}
\bigskip
Differentiating this functional gives
an equation for the local self energy and the nearest
neighbor self energy.
\GroupeEquations{
\begin{equation}
\Sigma _{loc} ={\frac{{\delta \Phi }}{{\delta G_{ii}}}}={\frac{{\delta
\Phi _{1}}}{{\delta G_{ii}}}}+z
\left(\,{\frac{{\delta \Phi _{2}}}{{\delta G_{ii}}}%
-\frac{{\delta \Phi _{1}}}{{\delta G_{ii}}}}
 \right) \\
\label{sigmalocal}
\end{equation}
\begin{equation}
\Sigma _{nn} ={\frac{{\delta \Phi _{2}}}{{\delta G_{ij}}}}
\label{sigmann}
\end{equation}}
The diagrammatic interpretation of the first equation is
transparent : each diagram that involves a link is properly
counted in the solution of a two impurity model.
The diagrams for the local self energy at site $i$ which involve
only the Green function   at site $i$, can be obtained from a one
impurity model with the correct counting.
The  contribution of the diagrams  for the local self energy which
involve the Green functions of a pair of sites,   can be
obtained by subtracting the contributions from the one and the
two impurity self energy and multiplying by $z$,
which results in equation  (\ref{sigmalocal}).
From $\Sigma _{loc}$ and  $ \Sigma _{nn}$ we can construct
the lattice self energy.
\begin{equation}
{\Sigma }^{latt}={\Sigma _{loc}(\iomn)+t(k)\Sigma _{nn}(%
\iomn)}
\end{equation}
and close the equations by requiring the self consistency condition
imposed by the Dyson equation \cite{DMFT}.
\GroupeEquations{
\begin{align}
G_{loc} &=\sum_{\ka}{\frac{{1}}{{\iomn-t(\ka)-\Sigma _{loc}(\iomn)-t(\ka%
)\Sigma _{nn}(\iomn)}}} \\
G_{nn} &=\sum_{\ka}{\frac{{e^{i\ka.\vec{\delta}}}}{{\iomn-t(\ka) -\Sigma
_{loc}(\iomn)-t(\ka)\Sigma _{nn}(\iomn)}}}
\end{align}
}
This can be expressed in the matrix notation, with $2\times 2$
matrices $G = \matrice{G_{loc} & G_{nn} \\ G_{nn} & G_{loc}}$ and $K$ in the
reduced Brillouin zone : 
\begin{equation}\label{Self.consis.PS.matrix.notation}
G  = 
\sum_{K \in R.B.Z.} \left( \iomn - t(K) - \sigmalatt (K)\right)^{-1}
\end{equation}
The generalisation of this scheme to antiferromagnetic order is
presented in Appendix \ref{App.Classical.Pair.Scheme}.
An important feature of this scheme is that the Green function of the one site problem
coincide with the diagonal part of the Green function of the two site
problem. This is a general ``nested'' structure for these schemes,
hence the name ``nested cluster schemes''. We will see in section
\ref{section.classicallimit} that this property leads to a quantitatively good
classical cluster scheme.


\subsection{Hartree-Fock terms}\label{sec.Hartree-Fock}

Another important issue is the treatment of longer range interactions within
CDMFT, and PCDMFT. In this context it is worth noticing that the Hartree
Fock contribution to the Baym Kadanoff functional (\ref{BKfunctional})
\begin{equation}
\Phi_{HF}[G]= \sum  U_{\alpha \beta \gamma \delta }(R_1, R_2, R_3, R_4)
 (G _{\beta \alpha}(R_2 R_1) G_{\delta\gamma  } (R_4
R_3 ) -G _{\delta \alpha}(R_4 R_1) G_{\beta \gamma  } (R_2 R_3)
)
\end{equation}
induces a self energy which is frequency independent and therefore
does not cause problems with causality and can be evaluated with
little computational cost. So it is convenient to separate $\Phi=
\Phi_{HF} + \Phi_{dyn}$,
and apply the cluster DMFT truncation only to $\Phi_{dyn}$, and
to the self energy it generates while treating the Hartree
contributions exactly. More precisely, one can treat with Hartree-Fock
terms that connect the cluster to the exterior only, to avoid a double
counting problem.

This observation is particularly relevant in the treatment of broken
symmetries induced by non local interactions as exemplified in the study of
the transition to a charge density wave in the extended Hubbard model in one
dimension which was studied in ref \cite{Kotliar-Venky-1d}.

\section{Classical limit}\label{section.classicallimit}

In this section, in an attempt to clarify the nature of the various cluster approximations,
we investigate analytically the large-$U$ limit  of the
Falikov-Kimball model, which reduces to the classical Ising model in
that case.
The Falikov-Kimbal model is defined by the Hamiltonian :
\begin{equation}\label{Falikov-Kimball}
H = \sum_{\langle i,j \rangle}
t_{ij\sigma} c_{i\sigma}^{\dagger}c_{i\sigma} + U
\bigl (n_{i\uparrow} - \frac{1}{2}\bigr ) 
\bigl (n_{i\downarrow}- \frac{1}{2}\bigr ) 
\end{equation}
where $\langle i,j \rangle$ denotes nearest neighbors, $t_{ij\uparrow} = t$ and $t_{ij\downarrow}=0$.
We consider the particle-hole symmetric case ($\mu =0$).
This model has been studied a lot for its own interest (for a review,
see \cite{Falikov-Kimball-Revue}), 
but we will use here as a tool to derive a classical
limit for the various cluster methods. This completes the QMC study of
this model with CMDFT and \DCA\  \cite{Jarrell-Falikov-Kimball}.
Indeed, in the  limit $U\rightarrow \infty$ with  $\beta \rightarrow
\infty$, $\bar \beta = \beta t /U$ fixed, it reduces on the lattice to the Ising model at
temperature $1/\bar \beta $ :
\def\jlatt{J^{\text{Latt}}}
\GroupeEquations{
\begin{align}
\label{Falikov-Kimbal-U-infty}
H &=  \sum_{\langle i,j \rangle} \jlatt_{i-j} S_{i}^{z}S_{j}^{z}
\\
\jlatt_{i-j} &=  t_{ij}/2
\end{align}
}
with  $S_{i}^{z}\equiv (c_{i\uparrow}^{\dagger}c_{i\uparrow} -
c_{i\downarrow}^{\dagger}c_{i\downarrow})$ (Ising spins). Note that a
factor $t/U$ has been absorbed in the definition of $\bar \beta $.
The proof is analogous to the standard reduction of the Hubbard model
to the Heisenberg model : since down electrons are quenched, there is
no possible exchange between two spins at different sites, which
implies that the interaction is Ising-like. 
We will now take the classical limit of the various cluster schemes,
using their common expression  (\ref{Eq.Cluster.2}) and
(\ref{ClusterEqGal.SelfConsistency})
in order to  obtain
classical cluster approximations of the Ising model.
For the reader not interested in the derivation,
we summarize our findings in section \ref{sec.class.discuss}. 

\subsection{Derivation of the classical limit}

We first present the derivation of the classical limit of
\CDMFT, \DCA\  and \PCDMFT. Our results on the classical limit of the Pair scheme are
presented but derived in appendix \ref{App.Classical.Pair.Scheme}.

In the following, we will concentrate ourselves on square clusters on 
the square lattice.
For clusters with an odd size
it is necessary to slightly generalize the \CDMFT\  equations to get
antiferromagnetism in the usual way, distinguishing two sublattices
\cite{DMFT}. To avoid cumbersome notations, we will discuss that point later.
The key point is of course that  $t_{ij\downarrow}= 0$ makes the model partially solvable in
the various schemes.

A priori, we have an effective action given by (\ref{Eq.Cluster.2}).
Since $t_{\downarrow}=0$, we can show self-consistently that
$G_{0\downarrow}$ is diagonal.
Indeed, if $G_{0\downarrow}$ is diagonal, because of the form of the
$n_{\uparrow}n_{\downarrow}$ vertex, $G_{\downarrow}$ is diagonal (at
all order in $U$), and so is $\Sigma_{\downarrow}$. In all schemes,
$\sigmalatt_{\downarrow}$ is then diagonal and independant of $K$. Therefore from
(\ref{Eq.Cluster.2}), we get for \CDMFT, \DCA\  and \PCDMFT :
\[
\bigl( G_{0\downarrow } \bigr )^{-1} (i \omega_{n}) =i \omega_{n} + 
\sigmacluster_{\downarrow }(i \omega_{n}) - \sigmalatt_{\downarrow
}(i \omega_{n})
=  i \omega_{n}
\]
(For the pair scheme, see Appendix \ref{App.Classical.Pair.Scheme}).
Therefore  we can consider $n_{\downarrow}$ as a classical variable
and compute the Green function
for the up electrons, solving the effective action :
\begin{equation}\label{class.effectiveaction}
  S_{\text{eff}} = - \iint_{0}^{\beta } d \tau  d \tau'
c^{\dagger}_{\mu  \uparrow} (\tau )
G_{0, \mu \nu  }^{-1} (\tau ,\tau')
c^{ }_{\nu  \uparrow} (\tau' )
+
\int_{0}^{\beta } d \tau
c^{\dagger}_{\mu  \downarrow} (\tau )
\partial_{\tau} 
c^{ }_{\mu  \downarrow} (\tau)
+ 
U
\Bigl (n_{\mu \uparrow}(\tau )  - \frac{1}{2}\Bigr ) 
\Bigl (n_{\mu \downarrow}(\tau )- \frac{1}{2}\Bigr )
\end{equation}
where $\mu ,\nu$ are cluster indices. For fixed $n_{\downarrow}$, the
action for the up electrons is Gaussian, which  leads to :
\GroupeEquations{
\begin{align}\label{Gup}
G_{\mu \nu } &= \sum_{ \{n_{\rho \downarrow}=0,1 \}} \frac{ Z ( \{n_{\rho \downarrow}\})}{Z}
\left(
G_{0\mu \nu }^{-1} - (n_{\mu \downarrow} - \frac{1}{2})U\delta_{\mu\nu }
 \right)^{-1}
\\\label{Gup2}
 Z ( \{n_{\rho \downarrow}\}) &\equiv  \exp \Biggl (
\Tr \ln \biggl (
G_{0\mu \nu }^{-1} - \bigl (n_{\mu \downarrow} - \frac{1}{2}\bigr) U\delta_{\mu\nu}
\biggr)
\Biggr)e^{\beta U/2
\sum_{\rho }n_{\rho \downarrow} }
\\
Z &\equiv \sum_{ \{n_{\rho \downarrow}=0,1 \}} 
 Z ( \{n_{\rho \downarrow}\}) 
\end{align}}

\def\omtilde{\tilde{\omega}}
The computation of the large-$U$ limit is organized in two steps : first, we find the
expansion of $G_{0}$, and second we show that in this limit the
effective action (\ref{Gup2}) becomes the action for
a classical Ising cluster with mean-field-like terms.  
Moreover, the study of the large-$U$ limit requires an expansion in the limit
$U\rightarrow \infty $ {\it and} $\omega \rightarrow \infty$, with
$x = 2\omega_{n} /U$ fixed. Indeed there is a strong $\omega$
dependence at the scale $U$, as seen already in the atomic limit.
We will see that the final result is {\it not} determined by the
$x\rightarrow 0$ limit.

First, we use an Ansatz for $G_{0}$, that we will prove to be
consistent 
\begin{equation}\label{class.Ansatz.G0}
G_{0}^{-1} (i\omega_{n})= i \frac{U}{2}x  -  \Delta (x) 
\end{equation}
with $\Delta$ of the order $1$ in $1/U$ (plus subdominant terms).
Using the  expansion  
\[
 \left ( \Lambda - \Delta  \right)^{-1} = \Lambda^{-1}
+ \Lambda^{-1} \Delta \Lambda^{-1} + \dotsb  
\]
with $ \Lambda_{\mu \nu } =  (S_{\mu } + i x)\frac{U}{2}\delta_{\mu \nu }$,
$S_{\mu }\equiv  1- 2 n_{\mu \downarrow}$, 
we expand $G_{\uparrow}$ in (\ref{Gup}) and obtain from (\ref{ClusterEqGal.DefSelf})
the following expansion for the cluster self-energy of the up
electrons $\sigmacluster$ : 
\GroupeEquations{\label{Classical.Sigma.From.Seff}
\begin{align}
\sigmacluster_{\mu\nu} &\equiv 
{\sigmacluster}^{\text{diag}}_{\mu \nu} + \delta \sigmacluster_{\mu \nu }\\
{\sigmacluster}^{\text{diag}}_{\mu \nu} &= -
\frac{U\delta_{\mu\nu}}{2}\left(\frac{1}{\moy{h_{\mu }}} - ix  \right)
+  O (1)\\
 \delta \sigmacluster_{\mu \nu } &\equiv ( 1- \delta_{\mu \nu})
\Delta_{\mu\nu}\frac{\moy{h_{\mu}h_{\nu}}_{c}}{\moy{h_{\mu}}\moy{h_{\nu}}} 
+ O \parent{\frac{1}{U}}
\\
h_{\mu } & \equiv \frac{1}{S_{\mu } + i x }
\end{align}}
where ${\sigmacluster}^{\text{diag}}$ (resp. $\delta
\sigmacluster_{\mu \nu }$) is the diagonal (resp. off-diagonal) part of the
matrix, the brackets denote the average over $n_{\mu\downarrow}$ or
$S_{\mu }$ with  the weights defined in (\ref{Gup}).
The averages and correlations of $h_{\mu }$ are computed using
$S_{\mu } = \pm 1$ and solving for the probability of the
spin to be $\pm 1$ as a  function of the correlations:
\GroupeEquations{
\begin{align}\label{class.relationshS}
\moy{h_{\mu }} &= \frac{\moy{S_{\mu }} - ix }{1 + x^{2}}\\
\moy{h_{\mu }h_{\nu }} &= \dfrac{\moy{S_{\mu }S_{\nu }} - ix \bigl (\moy{S_{\mu }} +
\moy{S_{\nu }}\bigr ) -  x^{2}}{\bigl (1+x^{2} \bigr )^{2}}
\end{align}}
For $(\mu ,\nu )$ nearest neighbors in an antiferromagnetic phase, as
well as in the paramagnetic phase, we have
$\moy{S_{\mu }}+\moy{S_{\nu }}=0$ and thus: 
\begin{equation}\label{class.relationshS.AF}
\dfrac{\moy{h_{\mu }h_{\nu }}_{c}}{\moy{h_{\mu }h_{\nu }}} =
\dfrac{\moy{S_{\mu }S_{\nu }}_{c}}{\moy{S_{\mu }S_{\nu }} -x^{2}}
\end{equation}
We now take the limit in the self-consistency condition
(\ref{ClusterEqGal.SelfConsistency}). We keep $\Sigma_{S}$ to derive a
general formula, and we will specialize to various schemes later.
We only use the fact that the diagonal part of $\Sigma_{S}$ is
independent of $K$ and equal to the cluster self-energy : 
$\Sigma_{S\mu \mu } = \sigmacluster_{\mu \mu} $. This is true for the
schemes studied in this paragraph : \PCDMFT, \CDMFT and \DCA (For the
pair scheme, see Appendix \ref{App.Classical.Pair.Scheme}).
Therefore, the dominant part of order $U$ in $\Sigma_{S}$ is diagonal
and we have :
\begin{equation}
\Sigma_{S} (K) = {\sigmacluster}^{\text{diag}}_{\mu \nu} +
\delta \Sigma_{S} (K) 
\end{equation}
where  $\delta\Sigma_{S}$ is a matrix of order $O (1)$ in the $1/U$ expansion.
Denoting with a bar the normalized sum of the reduced Brillouin zone,
\[
\overline{A (K)} \equiv \dfrac{\displaystyle \sum_{K \in \RBZ} A
(K)}{\displaystyle \sum_{K \in\RBZ} 1}
\]
we obtain ($R$ is on the superlattice) :
\GroupeEquations{\label{Classical.Expansion.Selfcons}
\begin{align}
\Delta_{\mu \nu } (x)&=
\tilde{t}_{\mu \nu } (x)+ \frac{2t}{U} J^{\mu \nu}_{\rho } (x)\moy{h_{\rho}}  + O \parent{\frac{1}{U^{2}}}
\\
\tilde{t} (x) &\equiv  \overline{t_{S}} +  \overline{\delta \Sigma_{S}}
- \delta \sigmacluster
\\
\label{Classical.Expansion.Selfcons.J}
 J^{\mu \nu}_{\rho } (x) &\equiv  
\frac{1}{t}
\left(
\overline{\bigl(t_{S} + \delta \Sigma_{S} \bigr)_{\mu \rho }
\bigl(t_{S} + \delta \Sigma_{S} \bigr)}_{\rho \nu }
- 
\overline{\bigl(t_{S} + \delta \Sigma_{S} \bigr)}_{\mu \rho }
\overline{\bigl(t_{S} + \delta \Sigma_{S} \bigr)}_{\rho \nu }
\right)
\\
\nonumber
&=
\frac{1}{t}
\sum_{R\neq 0} 
\bigl(t_{S} + \delta \Sigma_{S} \bigr)_{\mu \rho } (R)
\bigl(t_{S} + \delta \Sigma_{S} \bigr)_{\rho \nu } (-R)
\end{align}}
Note that $\tilde{t}$ is purely off-diagonal.
In the expression of $J$, $\delta  \Sigma_{S}$ has to be expanded to order $O(1)$
only. Moreover, only the off-diagonal part (in site index) is
important since the on site part is restricted to $R=0$.
$\delta \Sigma_{S}$ and $\Delta$ are determined by  Eqs. (\ref{Classical.Sigma.From.Seff},
\ref{Classical.Expansion.Selfcons}) and the relation between
$\Sigma_{S}$ and $\sigmacluster$.

At this stage, it is useful to distinguish two cases depending on the
validity of the cancellation 
\begin{equation}\label{cancellation}
\overline{\Sigma_{S}} \stackrel{?}{=} \sigmacluster
\end{equation}
In \CDMFT\  and \DCA, $\Sigma_{S}=\sigmacluster$ is $K-$independent, therefore
(\ref{cancellation}) holds and $\delta \Sigma_{S}$ drops out of
(\ref{Classical.Expansion.Selfcons.J}). 
$\tilde{t}$ and $J$ do not depend on $x$, although $\moy{h_{\mu }}$ does.
In \PCDMFT\  however, (\ref{cancellation}) does not hold and we have to
solve for $\delta \Sigma_{S}$ and $\Delta$ (See below) to complete the
computation of $G_{0}$.

The second step of the computation is to take the large-$U$ limit of
the effective action (\ref{class.effectiveaction}) using the value of $G_{0}$
(\ref{Classical.Expansion.Selfcons}). In that limit, we expect 
that the problem becomes classical,
and more precisely  for $U\rightarrow \infty $:
\GroupeEquations{\label{class.freezing}
\begin{align}
\moy{c_{\sigma \mu}^{\dagger} c_{\sigma \nu}} &\rightarrow 0 \qquad
\text{ for } \mu \neq \nu\\
n_{\mu \uparrow} + n_{\mu \downarrow}  &\rightarrow 1 \qquad \forall
\mu 
\end{align}}
Indeed this can be shown explicitly using (\ref{Gup}) and taking the
$U\rightarrow \infty$ limit in the Fermi factors, {\it after} doing
the summation over frequencies.
However due to the frequency dependent nature of $G_{0}$ in {\it all}
cases, we need 
to use the functional formalism and not the Hamiltonian formalism as in the derivation
of the Ising limit on the lattice. This is presented in detail in
Appendix \ref{App.details.class.Ugrand}.
In the limit $U\rightarrow \infty$ with  $\beta \rightarrow
\infty$, $\bar \beta \equiv \beta t/U$ fixed, the effective action  reduces to the
classical action for the cluster {\it at temperature} $1/\bar \beta $
and  we obtain a {\sl classical cluster scheme} for
the Ising model with an interaction $J_{\text{Ising}}$ inside the
cluster and an additional mean field term $J_{B}$ : 
\GroupeEquations{\label{Classical.cluster.scheme}
\begin{align}
H_{\text{eff}} &= \sum_{\moy{\mu ,\nu} } J_{\text{Ising}}^{\mu \nu } S_{\mu } S_{\nu}+ \sum_{(\mu ,\nu) }J_{B}^{\mu \nu }
S_{\mu}\moy{S_{\nu}}_{H_{\text{eff}}}
\\
\label{Classical.cluster.scheme.JIsing}
 J_{\text{Ising}}^{\mu \nu } &\equiv  \int_{-\infty}^{\infty}
\frac{ dx}{\pi t } \dfrac{ \tilde{t}^{2}_{\mu \nu } (x)}{(1+x^{2})^{2}}
\\
\label{Classical.cluster.scheme.JB}
 J_{B}^{\mu \nu } &\equiv  \int_{-\infty}^{\infty} \frac{ dx}
{\pi } \dfrac{J^{\mu \mu}_{\nu } (x)}{(1+x^{2})^{2}}
\end{align}
}
where $(\mu ,\nu)$ and $\moy{\mu ,\nu}$ correspond to a general and a
nearest neighbors couple of sites respectively. 
One has to solve a classical Ising cluster, with self-consistent
``boundary'' condition represented by $J_{B}$, which generalize the
usual Weiss field.
Of course, if $\tilde{t}$ does not depend on $x$, we find the same result as in
the lattice $J_{\text{Ising}}^{\mu \nu } = \jlatt_{\mu - \nu }$ for nearest neighbors.

Let us now specialize our computations to the three schemes presented
in Section \ref{sec.desc.methods} and compute the value of $J_{\text{Ising}}$ and $J_{B}$ for
\CDMFT, \DCA\  and \PCDMFT\  for square clusters of linear size $L$ on
a two-dimensional square lattice and for a general hopping $t_{\delta}$ where $\delta$
is a lattice vector.

\allowbreak 
\begin{itemize}
\item {\bf CDMFT} 
\item [] In this case, $\Sigma_{S} (K)= \sigmacluster$ is $K$
independent and $t_{S}=t$, which leads to $\tilde{t} (x) = t$ and :
\GroupeEquations{
\begin{align}
J_{\text{Ising}}^{\mu \nu } &= \jlatt_{\mu-\nu}\\
J_{B}^{\mu \nu } &= (-1)^{L}\sum_{R\neq 0} \jlatt_{\mu - \nu + R} 
\end{align}}
The interaction inside the cluster is the same as in the
lattice problem and the $J_{B}$ term is of order $O (1)$ and is confined to the boundary of
the cluster.
The boundary term couples a spin to the average value of its ``ghost''
neighbor in the neighboring cells, this average value being
computed in the cluster itself using the translation invariance on the
superlattice.
We added a $-$ sign for odd cluster size, since in this case, the
\CDMFT\  has to be generalized like \DMFT\  with two sublattices in order to
capture antiferromagnetism : this is equivalent to reversing the sign
of the ``ghost'' neighbor.
In the large cluster limit $L\rightarrow \infty$, the boundary terms
play no role and we therefore recover the lattice Ising problem.
Notice however that the one dimensional case is pathological
since the two boundary terms communicate with each other resulting in a finite $T_{c}$ in the limit
of infinite size. This pathology disappears in higher dimensions and explains the results
of reference \cite{Jarrell-Falikov-Kimball}.

\item {\bf  DCA}
\item [] In $d=2$, we use a square cluster of linear size $L=2L_{c}$, corresponding
to $L_{c}$ $K_{c}$ points (Cf Appendix \ref{AppendixDCA}). 
 With the definition $\sin_c (\vec{x})\equiv \prod_{i=1}^{d}\sin(x_{i})/x_{i}$ : 
\GroupeEquations{
\begin{align}
J_{\text{Ising}}^{\mu \nu } &= 
\frac{t}{2}
\parent{
\sum_{\empile{\delta }{\delta =\mu -\nu [2L_{c}]}}
\dfrac{t_{\delta}}{t}
\sin_c\parent{\frac{\pi \delta}{2L_{c}}}
}^{2}
\\
J_{B}^{\mu \nu } &= \frac{t}{2}
\sum_{\empile{\delta,\delta'} { \empile{{\delta =\mu -\nu [2L_{c}]}}{{\delta' =\mu -\nu [2L_{c}]}}}}
\dfrac{t_{\delta} t_{-\delta'}}{t^{2}} 
\sin_c\parent{\frac{\pi (\delta -\delta')}{2L_{c}}} 
-
J_{\text{Ising}}^{\mu \nu }
\end{align}}
$J_{B}$ is the same for all links as required by translation
invariance in the cluster.
For the first neighbor hopping,  denoting by $\jlatt$ the value of $\jlatt_{i-j}$ for $i,j$
nearest neighbors, the formula reduce to :
\GroupeEquations{
\begin{align}
J_{\text{Ising}}^{\mu \nu } &=  \frac{16 }{\pi^{2} } \jlatt
&
J_{B}^{\mu \nu } &= 
 \parent{2-\frac{16}{\pi^{2}}} \jlatt & \qquad \text{ for } L_{c}=1\\
J_{\text{Ising}}^{\mu \nu } &= \parent{\sin_c \frac{\pi }{2L_{c}} }^{2}\jlatt
&
J_{B}^{\mu \nu } &= 
 \parent{1- \parent{\sin_c \frac{\pi }{2L_{c}}}^{2}}
\jlatt &
\qquad \text{ for } L_{c}>1
\end{align}}
Note that in the $2\times 2$ cluster, the hopping is doubled
since electrons can hop from one site to the neighbor either directly or using the cyclicity condition.
In the large cluster limit $L_{c}\rightarrow \infty$, we recover the
lattice problem: $J_{\text{Ising}}^{\mu \nu} \rightarrow \jlatt_{\mu
- \nu }$ and $J_{B} = O (1/L_{c}^{2})$ \cite{Jarrell-Falikov-Kimball}.
\item {\bf PCDMFT}
\item [] 
This case is more complicated since $\Sigma_{S} = \sigmalatt$ is now
$K$ dependent, hence $\overline{\Sigma_{S}} \neq  \sigmacluster$,
{ except for diagonal elements.}
Therefore $t$ and $J$ now depend on $x$. In order to determine them,
we have to solve
Eqs.
(\ref{Def.SigmaLatt.AF},\ref{Classical.Sigma.From.Seff},\ref{Classical.Expansion.Selfcons}).
This is done in Appendix \ref{App.classical.calcul.pcdmft} and we
obtain : 
\GroupeEquations{\label{JIsingPCMDFT}
\begin{align}
\label{JIsingPCMDFT1}
\tilde{t}_{\mu \nu } (x) &=  t_{\mu \nu} 
\dfrac{\moy{h_{\mu}}\moy{h_{\nu}}}{A (x)\moy{h_{\mu}h_{\nu}}}
\\
\label{JIsingPCMDFT2}
J^{\mu \mu }_{\nu } (x) &=  (-1)^{L} A^{-2} (x) \sum_{R\neq 0} 
\dfrac{t^{2}_{\mu - \nu + R}}{t} 
\\\label{JIsingPCMDFT3}
A (x) &\equiv   1-\dfrac{1}{S_{c}}\sum_{{\empile{\rho \in
{\cal  C}}{\rho+\delta  \in {\cal  C}}}}
\dfrac{\moy{h_{\rho }h_{\rho +\delta }}_{c}}{\moy{h_{\rho }h_{\rho+\delta }}}
\end{align}}
where $\delta$ is one of the  basis vectors of the square lattice.

In $d=1$ for  first neighbor hopping, the last term reduces to
\[
A (x) = 1-\dfrac{1}{L}\sum_{i=1}^{L-1} \frac{\moy{h_{i}h_{i-1}}_{c}}{\moy{h_{i}h_{i-1}}}
\]
The boundary terms are located as in \CDMFT.
In the large cluster limit  $L\rightarrow \infty$, in
Eq. (\ref{JIsingPCMDFT}) the boundary terms
are subdominant in the denominator and the terms involving the
averages of $S$ cancel, restoring the correct $J_{\text{Ising}}$ on the lattice. The
$J_{B}$ term cancels since it is  restricted to  the boundary.
\item {\bf Pair scheme}
\item []
The computation of the classical limit of the pair scheme is a little
more involved but similar. 
We show in Appendix  \ref{App.Classical.Pair.Scheme}  that {\sl if a magnetic solution
exists}, it satisfies the equation of the classical variation method :
let's take two cluster problems, the first one with one site (denoted
with an index $(1)$) in a field $-z h$, the second one with two sites  (denoted
with an index $(2)$),
interacting antiferromagnetically with $\jlatt$ and in a field
$-(z-1)h$, where $z=4$ is the connectivity :
\GroupeEquations{\label{def.CVM}
\begin{align}
H^{(1)} &= z h S^{(1)} \\
H^{(2)} &= (z-1) h (S_{1}^{(2)} - S_{2}^{(2)}) + \jlatt S_{1}^{(2)} S_{2}^{(2)}
\end{align}}
where the field $h$ is determined by the ``nested consistency condition''
\begin{equation}\label{def.CVM2}
<S^{(1)}> = <S_{1}^{(2)}> = -<S_{2}^{(2)}>  
\end{equation}
Solving the classical equations, the critical temperature is given by : 
\begin{equation}\label{Critical.temp.CVM}
 (z-1) \tanh \bigl ( \bar \beta \jlatt \bigr) =1
\end{equation}
which gives $\bar T_{c}/\jlatt  \approx 2.88$.
\end{itemize}

\subsection{Discussion}\label{sec.class.discuss}

Let us recapitulate our results. We applied the cluster dynamical mean field
theories to the Falikov Kimball model taking $U\rightarrow \infty$ with  $\beta \rightarrow
\infty$, $\bar \beta = \beta t /U$ fixed. In this case the quantum model reduces to
the Ising model and the cluster dynamical mean field approximation to 
classical cluster methods.
\CDMFT\  corresponds to a simple mean field theory generalized to
clusters: each spin on the boundary is subjected to a mean field 
representing the interaction with the neighboring spins that do not belong to the cluster.
The form of the mean field is the standard one: the antiferromagnetic coupling times the 
magnetization of the spins. Note that in principle the antiferromagnetic coupling
is generated via quantum fluctuations and its value is
approximation-dependent. In the case of \CDMFT\  one gets the same
coupling as the one obtained for the lattice.

The classical limit of \DCA\  is different mainly for two reasons:
{\it i)}
the mean field is not just on the boundary but it acts on all sites and is
equal on sites belonging to the same sub-lattice. This is natural
in \DCA\  because it's an approximation that preserves the
translation invariance (or the reduced translation invariance for the AF phase),
in the propagator {\it and} in the Weiss field; 
{\it ii)} the form of the mean field is the standard one but the value of the
antiferromagnetic coupling in the mean field and for spins inside the
cluster are different from the lattice one (and between themselves).
Of course in the limit of infinite cluster they reduce to the lattice value.\\
\PCDMFT\  is  similar to \CDMFT\  to the extent that the mean
field acts only on the boundary. However, as for \DCA, the value of the
antiferromagnetic coupling in the mean field and for spins inside the
cluster are different from the lattice one (and between themselves).
Finally, the pair scheme reduces to the classical cluster variation
method for two sites.

The self-consistent equations corresponding to these
classical cluster schemes can be solved analytically for small
clusters and using a classical Monte Carlo method to solve the
impurity problem for larger size.
With the exception of the cluster variational method,
the classical limits of these extensions of \DMFT\  do not
result in drastic improvements in the estimation of $T_{c}$ for
small cluster sizes.
The value of $T_{c}$ predicted by DMFT is the standard mean
field one: $4J$ (the connectivity of the lattice is four) which is far
from the exact value that is $2.27J$.  The value obtained using the
NCS scheme with two sites, i.e. the pair scheme, lead to a good
improvement: $T_{c}=2.88J$.
Instead the results for \DCA\  and \CDMFT\ even for larger sizes (4 by
4, 6 by 6 and 8 by 8 clusters) do
not improve the estimate of $T_{c}$ very much. In fact within error
bars of $0.1$ we get the $T_{c}^{(4\times 4)}=3.2J$, $T_{c}^{(6\times
6)}=3.J$ and $T_{c}^{(8\times 8)}=2.85J$ (within the error bars the
estimate for $T_{c}$ are the same for the two schemes). Note that the
value obtained with a cluster of $16$ sites with \DCA\  and \CDMFT\
reaches the estimate obtained with a cluster of two sites using NCS!
The results of \PCDMFT\  are clearly worse than the ones of the two
previous methods
for the numerical value of $T_{c}$ which is quite larger than what
is obtained by \CDMFT\  and \DCA.
This can be traced to the lack of cancellation of off-diagonal
elements in (\ref{cancellation})
which implies that the cavity field $\Delta(i \omega_{n})$ in \PCDMFT\ is
not proportional to the square of the
hopping matrix element which is an  undesirable feature.

\section{Causality}\label{section.causality}
In this section, we present a general method to prove the causality of
cluster approximations and apply it to various schemes, including
\PCDMFT. 
As mentioned in section \ref{sec.desc.methods}, there are two equivalent presentations for
the equations of \DMFT\  and its extensions : a first one using the Weiss
function $G_{0}$, and another one using the Luttinger-Ward functional
$\Phi$. Let's examine the causality question in both.
\begin{itemize}
\item In the formulation with $G_{0}$, one has to prove that: {\it i)} If $G_{0}$ is causal, $G_{c}$ and $\Sigma$ are causal  : this
is a automatic (as long as one uses a causal impurity solver).
{\it ii)} If $\Sigma$ is causal, $G_{0}$ computed from the self
consistency condition is causal. This is the difficult part. It was
carried out explicitly for \CDMFT\  in \cite{CDMFT} and for \DCA\  in
\cite{DCA-all}.
\item In the formulation with $\Phi$, one has to prove that: 
{\it i)} Given a causal $G$, $\Sigma_{skel} (G) = \delta \Phi/ \delta G$ is causal; {\it ii)}
Given a causal $\Sigma$, the self consistency condition produces a
causal $G$. Here {\it ii)} is obvious, because of the simple form of
the bare propagator. {\it i)} is the difficult part and this section
is devoted to a general method to prove it.
\end{itemize}

To show that the self-energy is causal we have 
to prove that the retarded self-energy has a negative imaginary part
for all $\omega$, {\it i.e.} 
$\Im \Sigma_{R} (x,x',\omega) \equiv  (\Sigma_{R} (x,x',\omega )-\Sigma_{R}^{*}(x',x,\omega ))/2i $ is
negative.
We use the Cutkovsky-t'Hooft-Veltman equation (also known in the literature as ``cutting
equations'') which is extremely useful to discuss causality properties in terms
of a diagrammatic expansion. It relates the imaginary part of the
self-energy to a sum of cut diagrams.
A standard derivation can be found in \cite{VeltmanBook} and it is 
due to t'Hooft and Veltman. However, we present in Appendix
\ref{AppendixKeldysh} a simpler and self contained derivation, based on the Keldysh 
method. Contrary to the previous derivation it does not assume
translation invariance, which is important to discuss
cluster schemes that break translation invariance.
Note however, that this method is limited to zero temperature formalism.

This section is organized as follows : 
in paragraph \ref{LT}, we present the Cutkovsky-t'Hooft-Veltman
equations
and show how it can used to prove causality; in paragraph
\ref{cutting.appl}, we apply it to various cluster schemes.

\subsection{The Cutkovsky-t'Hooft-Veltman equation}\label{LT}

The Cutkovsky-t'Hooft-Veltman equation gives the imaginary part of the
retarded self-energy as a sum of cut diagrams. Let us consider the set of all perturbative
connected diagrams $\{D \}$ for the self-energy at zero temperature
(for a given approximation) and all
possible cuts of these diagrams into two connected parts $L$ and
$R$ containing respectively the left and the right external point (if
they are the same, we drop the diagram).
There are $n$  cut propagators going from left to right and $n-1$
from right to left, at frequency $\omega_{i}$ and indices $x_{i}$. We denote by $MR$
the  mirror of $R$ defined as the left diagram obtained from $R$ by reversing all arrows.
(Cf. Fig. \ref{fig5.eps} for an illustration).
\begin{figure}[hbt]
\[
\figx{8cm}{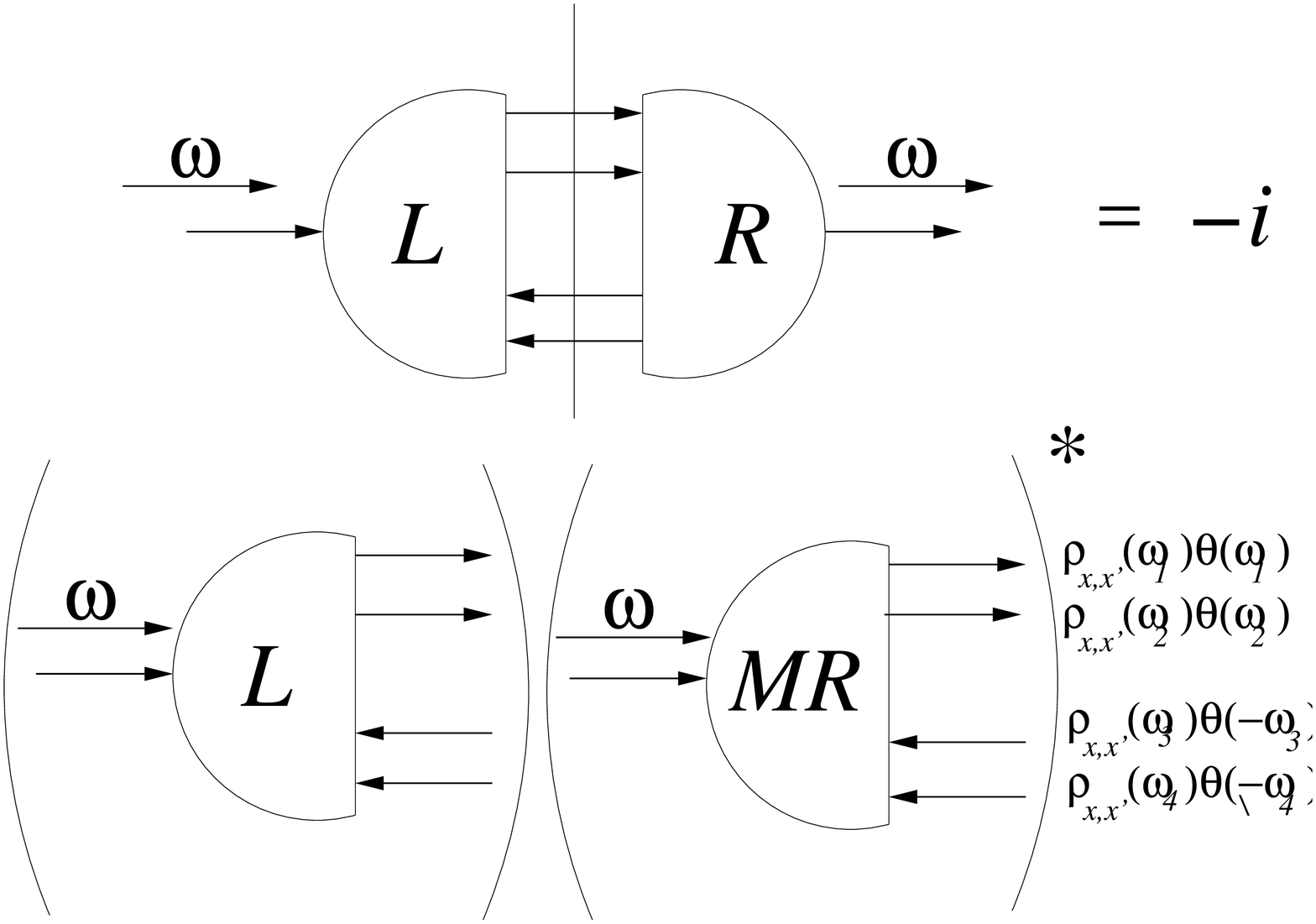}
\]
\caption{Definition of cut diagrams} 
\label{fig5.eps}
\end{figure}
To each left diagram, we associate the value $d_{L}$ of corresponding $T=0$
diagram, using the $T=0$ Feynman propagator ($G^{++}$ in Keldysh notation : see appendix
\ref{AppendixKeldysh}). Note that for each given half left diagram one has
to put in $d_{L}$ all the half left diagrams obtained permuting
the cut lines in all the possible ways and then multiply by an appropriate symmetry
factor discussed in the appendix (\ref{AppendixKeldysh}).
We then write 
${D}_{L} (x,\omega ,\{ x_{i}\},\{\omega_{i}\}) \equiv  d_{L} \theta (\omega) +
d_{L}^{*} \theta (-\omega)$,  where $\theta$ is the Heaviside function.
We denote all cut propagators with an index $i$.
To each we associate
$\rho (x_{i},x_{i}',\omega_{i})\theta (\epsilon_{i} \omega_{i} )$ with
$\epsilon_{i} = \sgn \omega $ for
a propagator going to the right, $-\sgn \omega$ for a propagator going to the
left.
The summation over diagrams can be expressed as a summation over
$n$ and a sum over diagrams $D_{n} =({L_{n}},{R_{n}})$ 
with a given number of cut propagators going to the right.
The Cutkovsky-t'Hooft-Veltman equation is then:
\begin{multline}\label{Cutkovsky.eq}
\Im \Sigma_{R}(x,x',\omega)=
-
\sum_{n \geq 2}\frac{1}{2n!} \sum_{ D_{n}= (L_{n},R_{n})}
\sum_{\empile{\{x_{i} \}}{\{x'_{i} \}}}\int d\omega_{i}
{D}_{L_{n}} (x,\omega,\{x_{i} \},\{\omega_{i} \}) 
\bigl ({D}_{MR_{n}}(x',\omega,\{x_{i}'\},\{\omega_{i}\})\bigr
)^{*}\times \\
\times \prod_{i=1}^{n_{L}+n_{R}} \rho(x_{i},x_{i}',\omega_{i})\theta
(\epsilon_{i} \omega_{i} )
\end{multline}
{\it Remark:} if the system is translation invariant then we can Fourier transform 
also with respect to space and get:
\begin{equation}\label{Cutkovsky.eq.k}
\Im \Sigma_{R} (k,\omega )=
-
\sum_{n \geq 2}\frac{1}{2n!} \sum_{ D_{n}= (L_{n},R_{n})}
\sum_{\empile{\{x_{i} \}}{\{x'_{i} \}}}
 \int d\omega_{i}
{D}_{L_{n}} (k,\omega,\{k_{i} \},\{\omega_{i} \}) 
\bigl ({D}_{MR_{n}}(k,\omega,\{k_{i}\},\{\omega_{i}\})\bigr )^{*}
\prod_{i=1}^{n_{L}+n_{R}} \rho(k_{i},\omega_{i})\theta (\epsilon_{i} \omega_{i} )
\end{equation}
Note that there are no terms with $n \leq 2$, otherwise the original closed diagram would not be 2PI.

Let us now see how to use (\ref{Cutkovsky.eq}) to prove causality of
the self-energy. We want to show that:
\begin{equation}\label{th}
{\cal  A}=\frac{1}{2i}\sum_{x,x'}w_{x}\left(\Sigma_{R} (x,x',\omega )-\Sigma_{R}^{*}(x',x,\omega ) \right)w_{x'}^{*}
\le 0\qquad \forall \{ w_{x} \}
\end{equation}
for all complex vector $w_{x}$.
Using eq. (\ref{Cutkovsky.eq}) we find that we can write the previous quantity as a sum over cut diagrams:
\begin{multline}\label{Causality.proof1}
{\cal  A}=
-\sum_{\empile{n_{L}\geq 0}{n_{R}\geq 0}}\frac{1}{2n!}
\sum_{ D_{n}= (L_{n},R_{n})}
\sum_{\empile{\{x_{i} \}}{\{x'_{i} \}}}
 \int d\omega_{i}
\biggl(
\sum_{x} w_{x} 
{D}_{L_{n}} (x,\omega,\{x_{i}\},\{\omega_{i}\} )
\biggr) 
\biggl(
\sum_{x' } w_{x'}^{*}
{D}_{MR_{n}}^{*}(x',\omega,\{x_{i}'\},\{\omega_{i}\} )
\biggr) \times \\
\prod_{i=1}^{n_{L}+n_{R}} \rho(x_{i},x_{i}',\omega_{i})\theta (\epsilon_{i} \omega_{i} )
\end{multline}

At this stage, we can use the causality of $G$, {\it i.e.} the
positivity of $\rho (x,x')$. It implies the positivity of the tensor
product $\prod_{i=1}^{n_{L}+n_{R}} \rho(x_{i},x_{i}',\omega_{i})$
considered as a matrix of indices $(x_{1},\dotsc ,x_{n_{L}+n_{R}})$.
Hence, positivity of $-\Im \Sigma$ would result if the summation over
diagrams could be recast as a square.
More precisely, a simple property that implies causality is
that for any $n$ :
\begin{equation}\label{RequirementCausality}
\{({L_{n}},{R_{n}}) \} = \{{L_{n}} \} \times \{{R_{n}} \}
\end{equation}
$\{{L_{n}} \}$ (resp. $\{{R_{n}} \}$ ) is the set of left (resp.  right) parts of
diagrams with fixed $n$.
Eq. (\ref{RequirementCausality}) says that all diagrams can be obtained once and only once using 
a left part and a right part, which is equivalent to two properties : 
\begin{itemize}
\item Closure property : With any left part and any right part (that
have the same $n$), gluing them constructs a
diagram present in the expansion.
\item Exact counting property : all diagram are obtained that way
(by definition of left/right parts) but only once : there is no
overcounting problem
\end{itemize}
If the property (\ref{RequirementCausality}) holds, then using the
notation 
\[
y_{n} (\{x_{i} \};\{\omega_{i} \},\omega ) \equiv 
\sum_{x,L_{n}} w_{x} 
{D}_{L_{n}} (x,\omega,\{x_{i} \},\{\omega_{i} \} ) 
\]
we have :
\begin{equation}\label{Causality.proof2}
{\cal  A}=-\int d\omega_{i}
\sum_{n\geq 2}
\sum_{\{x_{i}\},\{x_{i}' \}}
y_{n} (\{x_{i}\};\{\omega_{i} \},\omega )
y_{n}^{*} (\{x_{i}'\};\{\omega_{i} \},\omega )
\prod_{i=1}^{n_{L}+n_{R}} \rho(x_{i},x_{i}',\omega_{i})\theta (\epsilon_{i} \omega_{i} )
\end{equation}
Note that the sum over $\{x_{i}\},\{x_{i}' \}$ inside the integral can
be interpreted as the product of the two vectors $y$ and $y^{*}$ with
a positive definite matrix. Thus the positivity of $\rho$ implies the negativity of $\Im \Sigma$.

The efficiency of this method is that one can prove causality,
{\sl just by examining the combinatoric structure of the diagrams
present in the approximation}, with no further computation.
A few remarks are important at this stage : {\it i)}
This method can prove the causality of a scheme, but it can not
prove that a scheme will violate causality. In particular it proves
the positivity of $-\Im \Sigma$ {\it for all causal propagators} $G$.
In the following, we will refer to this as the strong causality property.
It is a logical possibility that this property does not hold, while
the {\it actual solution of the scheme} $G$ is indeed causal for all
parameters. However, we know no example of this, and the strong
property is interesting since we use iterative algorithm to solve the self-consistent
impurity problem.
{\it ii)} Equation (\ref{RequirementCausality}) is sufficient but not necessary
: one could also have form squares in a more complicated manner.
{\it iii)} As a remark, we see that  the full perturbation theory on the lattice
satisfies the property (\ref{RequirementCausality}) : the full
diagrammatic expansion obviously leads to a causal self-energy.

\subsection{Applications}\label{cutting.appl}

\subsubsection{\CDMFT\  and \DCA}

\CDMFT\  can be obtained from the Baym-Kadanoff functional taking 
$\Phi=\sum_{\bf R}\Phi(G_{\alpha ,\beta ;\bf R,\bf R})$, where $R$ is the cluster index 
and $\alpha $ is the internal cluster index. 
As a consequence $\Sigma _{\alpha ,\beta ;\bf R,\bf R}=
\Sigma _{\alpha ,\beta }=\frac{\delta  \Phi}{\delta G_{\alpha ,\beta
}}$.
Since $\Phi(G_{\alpha ,\beta })$ is  the sum of all the 2PI diagrams
of a $L^{d}$ lattice, \CDMFT\   does satisfy
(\ref{RequirementCausality}) and therefore is a causal scheme.

In \DCA,  one writes the self energy in terms of the
complete series of diagrams and the delta function at the vertices are replaced by a
coarse grained function. As discussed previously \DCA\  is equivalent
to replace the Kronecker delta $\delta _{k_{1}+k_{2},k_{3}+k_{4}}$ with a Kronecker 
delta  $\delta _{k_{1}^{c}+k_{2}^{c},k_{3}^{c}+k_{4}^{c}}$ where $k^{c}_{i}$ is the cluster
momentum related to the hypercube containing $k_{i}$. Formally
$k^{c}_{i}=k_{i} \quad \mbox{mod}\quad (\pi /L,\dots ,\pi
/L)$.
Hence, \DCA\  satisfies (\ref{RequirementCausality}) : the proof is the same that for the complete series of
diagrams for the self energy. The fact that the $k$ structure of the vertices is different
does not change nothing at all. Hence, we see that \DCA\  is
automatically causal.

\subsubsection{PCDMFT}

The causality of \PCDMFT\  is easily established using its real space
formulation (\ref{Def.SigmaLatt.AF}).
The causality for \CDMFT\  implies the causality of $\sigmacluster$.
We just have to show that $\sigmalatt$ is causal.
For any complex vector $x_{\sigma\mu }$,
we have from (\ref{Def.SigmaLatt.AF})
\GroupeEquations{
\begin{align}
\sum_{\sigma \sigma' \mu \nu }
x_{\sigma\mu }^{*} x_{\sigma' \nu } 
\sigmalatt_{ \sigma, \mu ;\sigma', \nu } (K)
&= \frac{1}{S_{c}}
\sum_{ { \delta }\in \{0,\dotsc ,L-1 \}^{d}}
\sum_{\sigma \sigma' \bar \mu \bar \nu }
z_{\sigma ;\bar \mu}^{*} (\delta)
z_{\sigma'; \bar \nu}    ( \delta)
{\sigmacluster}^{\alpha ({\delta })}_{\sigma,\bar \mu ;\sigma' \bar \nu 
}
\\
z_{\sigma; \overline{\mu  + \delta }} (\delta) & \equiv  x_{\sigma \mu} 
e^{ i K \cdot \lfloor \frac{{\mu} + \delta}{\sizecluster}
\rfloor L}
\end{align}}
The negativity of $\sigmacluster$ implies that we have a sum of
negative terms. This shows that causality of PCMDFT 
is guaranteed {\sl term by term in the $\delta$ sum}.\\
Note that another way to reach the same conclusion is based on the
fact that \PCDMFT\  can be interpreted as the solution of a
real lattice systems replacing the real $U( k_{1},k_{2},k_{3},k_{4})$
with its \PCDMFT\  counterpart. Thus, the statement of causality 
of \PCDMFT\  is a particular case of the causality of the original
problem with a general $U$.

\subsubsection{Non-Causality of the pair scheme}\label{KG}
The pair scheme, described in Section \ref{sec.nested.clusters},
has been originally introduced in \cite{Ingersent-Schiller,DMFT}. 
It seems a very natural way to introduce systematically a k-dependence
for the self-energy. Unfortunately, as we will show in the following,
this scheme is not causal for two reasons that we shall discuss
in detail.

First, let's apply the Cutkovsky rule. We see that the pair scheme
does not satisfy the property (\ref{RequirementCausality}) or more
precisely the closure property.
Indeed, when cutting a diagram for $\Sigma$, we only have two sites
indices $i$ and $j$. Let consider a cut diagram $({L1},{R1})$ where we have $i$
along the cut, and the similar diagram $({L2},{R2})$ where $j$ is
replaced by $k\neq i,j$. A priori, we could glue the parts into
$({L1},{R2})$, but that diagram would involve $i,j$ and $k$, and
thus is not present in the diagrammatic expansion generated by the
pair scheme.
This is a first reason for lack of causality. We stress that this is 
not a proof, but a quick and strong indication that the scheme is not
strongly causal. It's interesting to point out that this problem was
already encountered in the case of disordered electron systems when 
people tried to generalize CPA to clusters. In that case Mills and 
collaborators \cite{Mills} noticed that a generalization of CPA similar 
to the pair scheme has this type of non-closure problem. For this reason they introduced the
traveling cluster method which is fully causal. We will comment 
afterward on the possibility to do the same for strongly correlated
electrons.

Let us now really show that the pair scheme is not causal. In fact 
the violation of the closure property is just an hint 
that there could be some problem with causality but it doesn't 
preclude the possibility that eventually the scheme is causal.
In the following we will consider the solution of the pair scheme 
when $U\rightarrow 0$ for an arbitrary bare spectral density or 
hopping on the original lattice. We will show that a particular choice
of the bare spectral density gives rise to a non causal self-energy
thus proving in an explicit example the non causality of the scheme. 

To study the sign of the imaginary part of the self-energy in the
limit $U\rightarrow 0$ one can focus just on the second order term
because the first order gives no contribution. Moreover one can 
replace in the diagram the full propagator with the bare or lattice one
($G_{bare}= (i\omega_{n}-t-\mu )^{-1}$), the error leading to a term
of an order larger than $O (U^{2})$.
If one applies the Cutkovsky equations to obtain the imaginary part of
the self-energy for the second order diagrams of the pair scheme then one has to restrict 
the sum in the first line of Fig. \ref{fig6.fig} to single sites $(x,x)$ and links $<x,x'>$. The sum over the single
sites $(x,x)$ is just the same that one obtains for usual \DMFT\  and is clearly positive. 
The link term $<x,x'>$ with $x,x'$ nearest neighbors equals ($\omega $ is the (positive) self-energy frequency):
\begin{eqnarray}\label{neg}
U^{2}\int_{0}^{+\infty } d\omega _{1}\int^{0}_{-\infty }&& d\omega _{2}\sum_{<x,x'>}\left(w_{x}w_{x'}^{*}
\rho (x,x',\omega _{1})\rho (x,x',\omega _{2})\rho (x,x',\omega -\omega _{2}-\omega _{3})+\right.\nonumber
\\
 &+&\left.w_{x'}w_{x}^{*}
\rho (x',x,\omega _{1})\rho (x',x,\omega _{2})\rho (x',x,\omega -\omega _{2}-\omega _{3})\right)
\end{eqnarray}
The spectral density in the previous equations is just the bare spectral density obtained by $t_{ij}$.
For the pair scheme one can find a spectral density matrix that, injected into the previous equations, does not lead to 
a causal self-energy. 
The causality of the spectral density matrix is equivalent to the positivity of each Fourier
element $\rho _{k}(\omega )\ge 0$. 
However, even if $\rho _{k}(\omega )\ge 0$, $\rho _{x,x'}(\omega )$
with $x,x'$ nearest neighbors can be negative and as close as possible (in absolute value) to $\rho _{x,x}(\omega )$
for any $\omega $.
Consider for example the example in which $\rho _{k}(\omega )$ is
strongly peaked and has an important weight only near the points $\pm
\pi,\dots ,\pm \pi  $. 
 In this case $\rho _{x,x}(\omega )>0$, $\rho _{x,x'}(\omega )$
with $x,x'$ nearest neighbors is negative and $\rho _{x,x'}(\omega
)=\int_{BZ}\frac{dk}{(2\pi)^{d}}\rho _{k}(\omega )\exp
(-ik(x-x'))\simeq -\int_{BZ}\frac{dk}{(2\pi)^{d}}\rho _{k}(\omega
)=-\rho _{x,x}(\omega )$. 
In this case the contribution of (\ref{neg}), for example for $w_{x}=1$, is negative and 
approximatively two times larger in absolute value than the contribution from the $x,x$ term.
Since for each $x,x$ term there are $z$ ($z$ is the connectivity) terms like (\ref{neg}) the global sum
is negative for $z>2$, i.e. in any dimension large than one (if one takes a square or cubic or hypercubic lattice). 
Thus in this case, the self-energy is not causal. 
So we have found a second reason for non-causality: even if
the closure property is verified and in the cutting-folding
procedure no new diagrams are created (which is the case for the pair
scheme at second order in $U$) this does not 
guaranteed that there is a way to write all the cut diagrams as a sum
of squares. One can have an overcounting problem as we have just shown
for the pair scheme.

\begin{figure}[bt]
\[
       \figx{8cm}{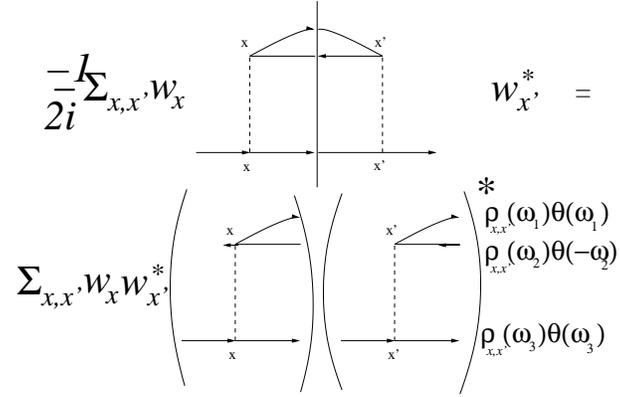}
\]
\caption{Example of cut diagram. MR means mirror image of R.}
\label{fig6.fig}
\end{figure}

Let us now comment on the possibility to cure the pair scheme as
Mills and collaborators did in the case of disordered electrons
defining the traveling cluster scheme. What they did is to solve the
violation of the closure property allowing an arbitrary self
energy diagram to have only single site and nearest neighbor
propagators. So in this case a self-energy diagram can connect two
arbitrary points but just through a sequence of nearest neighbor
propagators (that is why the scheme is called ``traveling'').
This clearly solve the first problem: the non-closure but 
in principle it doesn't guarantee that the second one also is
cured. And indeed it is not for strongly correlated electrons.
A simple way to see it is that even with a generalization of this 
type in the limit $U\rightarrow 0$ the causality counter-example 
exhibited for the pair scheme is still valid. Instead the case of
disordered electrons has a simpler diagrammatics (there are no loops
in the self-energy diagrams due to the use of the replica method) and one can explicitly
rewrite everything as a sum of squares. In particular the diagram 
discussed before is simply not there.
 
A way to circumvent these two difficulties is write the self-energy as the sum of all diagram and then
replace the original propagator with a simplified restricted version that is guaranteed to be causal.
For example in order to have a causal generalization of the pair scheme, close in spirit to
the traveling cluster, one can write:
\[
\Sigma =\left.\frac{\delta \Phi }{\delta G}\right|_{G=\tilde{G}}\qquad
\tilde{G}(k,\omega )=G_{0}(\omega )+G_{1}(\omega )\sum_{i=1}^{d}(2\cos
(k_{i})) 
\]
where $G_{0},G_{1}$ are fixed imposing the conditions
$\tilde{G}(0,\dots ,0,\omega )=G(0,\dots ,0,\omega )$ and
$\tilde{G}(\pi ,\dots ,\pi ,\omega )=G(\pi ,\dots ,\pi ,\omega )$, and  
$G=(G_{0}^{-1}-\Sigma )^{-1}$ This guarantees that $\tilde{G}$ is a causal propagator as long as 
$G$ is causal. So when $\tilde{G}$ is inserted in the complete series it 
will give rise to a causal self-energy. 
The problems with this scheme is that: (1) it seems that it is not $\Phi $-derivable.
The reason is that in all the diagrams there are only nearest-neighbor propagator. However
since the diagrams connect in general two sites that are not nearest neighbor it is difficult to
think how one can obtain it using a restricted set of diagrams for
$\Phi $. Note however that is also the case of the traveling cluster
scheme for disordered electrons.
(2) It is not clear what type of impurity solver, if any, can be used to sum the series of diagrams.
In the case of the traveling cluster many simplifications arise that
allow one to solve (2) but not (1).
the smaller becomes the causality violation.
Let us finally remark that the two mechanisms behind the non causality
of  the pair scheme apply also to the  general NCS schemes even though we
expect smaller violations of causality for larger clusters.

\section{Conclusion}
Cluster schemes are a promising method for studying strongly
correlated electrons. Compared to the study of small finite size
systems, they offer the advantage that the thermodynamical limit
is taking from the outset, and hence one can hope for smaller
finite size effects. As a prime  example,  we now know that even a
one site cluster captures many non trivial features of the Mott
transition.

On the other hand, there is no unique generalization of the single site 
dynamical
mean field approximation. Some generalizations view the
\DMFT\  and its resulting impurity (or multiple impurity models in
the case of the NCS) as a trick for summming selected classes of
diagrams. An alternative view is provided by
the cavity construction and by functional methods,  where
an effective action  is constructed to compute the correlation
functions of  selected coarsed grained or cluster degrees
of freedom. Finally, one can  view \DMFT\  as a choice of boundary
conditions for finite size studies of a small set of degrees
of freedom, where the boundary conditions recognizes that
those degrees of freedom are periodically repeated in a medium.
While all those points of view, lead to the same set of \DMFT\  equations
in the single site case, they lead to different constructions 
for cluster extensions,
and the strenghts and limitations of these  extensions need to be
explored. This paper contributed in this direction by clarifying
two important aspects of these extensions, namely the conditions on
the cluster scheme needed for the scheme to satisfy causality,
and the reduction of the cluster schemes to spin cluster methods
in the classical limit.

First, we introduced \PCDMFT\ 
which is a causal generalization of the scheme proposed by 
Lichtenstein and Katsnelson \cite{Lichtenstein-Katsnelson}.
Second, we provided a general way to prove causality of cluster scheme
and showed that the pair scheme, which  is the most natural extension in the 
sense that
it is defined by a translational invariant restriction of the
Baym Kadanoff functional, is not causal : the diagrammatic
reason for the failure of this method was clarified.
Third, we also showed formally, in the context of the Falikov Kimball
model how the semiclassical limit of the different cluster
methods reduce to classical spin cluster methods.
 Both \DCA\ and
\CDMFT\  give comparable answers even though their Weiss fields have a very
different form (in \DCA\ the Weiss field is uniformly distributed
inside the cluster while in \CDMFT\ it is focused on the boundary). 
On the other hand \PCDMFT, have an unphysical feature
in the classical limit which results from a Weiss field which
does not vanish as the square of the hopping matrix elements.
The nested cluster schemes are clearly superior and provide
a quantum generalization of the cluster variation method
which gives critical temperatures that converge
rapidly with cluster size. 
We showed that the nested cluster schemes violates causality. Our proof suggests  that the
this violation is going to be more severe, when  a substancial part of the
lattice self energy has a range that
exceeds the size of the maximal cluster used in the NCS.  On the contrary when the range of the self energy
is contained in the maximal cluster, the diagrams generated by the Cutkovsky procedure which
are not contained in the NCS are small, which is consistent with the early numerical studies which indicated
no violation of causality in the pair scheme at high temperature. This observation
is in the same spirit as a   recent study of  a  new
(non-nested) scheme, the ``fictive impurity model method'' , which   connected violation of causality with peaks in momentum
space in the lattice self energy \cite{Millis_fictif}.

An important lesson that can be learned from our results is that 
the choice of an optimal cluster scheme may depend
of the property studied and the physical system at hand.
Indeed our semiclassical analysis revealed \PCDMFT\ is
not accurate in insulating regimes, at least in the one that we
focused on doing the semiclassical limit. Instead \CDMFT\ performs
rather well. We conjecture that \CDMFT\ is accurate 
in insulating cases and \PCDMFT\ in metallic cases.
Indeed a recent numerical in one dimension  work points in this direction 
(\cite{CaponeCivelli}).

Another important result concerns the remarkable accuracy
of the NCS scheme compared to the other ones. Even if its performance
in a case with stronger quantum fluctuations remains 
an open question, we think that it would particulary interesting 
to try to cure the causality problem keeping the nesting idea 
inherent to the NCS scheme.
Another route to follow is try to incorporate the cellular dynamical
mean field ideas of defining impurity  models in adaptive basis sets
into a NCS scheme. In this way one could try to adapt the basis
to the problem so that the resulting self energy is short range and
the causality problem, even if not avoided, is sensibly reduced.  

Finally, while the paper focused on  a few schemes the techniques developed here are 
quite general and
may play an important role in selecting and optimizing cluster methods for
specific applications.

\acknowledgments 
We thank A. Lichtenstein, A. Georges, S. Florens, S. Biermann for
useful discussions, and M. Jarrell for pointing to us Ref. \cite{MaierPhD}.
This work was supported by  NSF grant N$^{\circ}$ DMR -0096462.
We acknowledge the warm hospitality of the KITP program on Realistic
Studies of Correlated Electrons,  at UCSB (NSF grant PHY 99-07949) where
part of this work was carried out, and the  participants of this workshop
for discussions.

\appendix

\section{Real space formulation of \DCA}\label{AppendixDCA}
In this appendix, we present a real space formulation of \DCA
\cite{Kotliar-Biroli}, which is useful for orders that break
translation invariance.
Let us consider first the translation invariant case.
The hopping of the full lattice can be written using the reciprocal
superlattice and the cluster indices (in real space) : 
\begin{equation}\label{App.DCA.t.supernotations}
t_{\mu\nu} (K) = \frac{1}{L^{d}} \sum_{K_{c}} e^{i (K+K_{c}) \cdot (x_{\mu} -x_{\nu}
)} t (K+K_{c})
\end{equation}
Here (in d=1), $-\frac{\pi}{L} \leq K \leq \frac{\pi}{L}$, $K_{c}=
\frac{2\pi n}{L}, n=0,\dotsc,L-1$ is on the reciprocal lattice of the
(finite) cluster.
Using $t_{S}= t$ in (\ref{ClusterEqGal.SelfConsistency}), we get \CDMFT. Using $t_{S}=t^{cyc}$ with 
\GroupeEquations{\begin{align}\label{def.tcyc}
t_{\mu\nu}^{cyc}(K) &\equiv  \frac{1}{L^{d}} \sum_{K_{c}} e^{i K_{c} \cdot (x_{\mu} -x_{\nu}
)} t (K+K_{c})\\
& = e^{-iK x_{\mu} }t_{\mu\nu} (K) e^{iK x_{\nu}} 
\end{align}}
we get \DCA. Indeed, $t^{cyc}$ is cyclic in the cluster indices (by
definition of $K_{c}$), and {\sl provided that we obtain a translation
invariant (cyclic) solution} we can diagonalize all matrices in the cluster to get
Eq. (\ref{def.DCA}).

Let us now consider the case where the solution of the real space
formulation of \DCA\  breaks the
translation invariance in the cluster into a smaller invariance : the
(big) cluster is divided into $L_{c}^{d}$ small clusters of linear size $L$, and we have
translation invariance in the big cluster  when the small clusters are
thought as collapsed into  one point.
For example, for AF order in $d=1$, $L=2$ (See Figure \ref{emboitement.eps}).
\begin{figure}[hbt]
\[
       \figx{8cm}{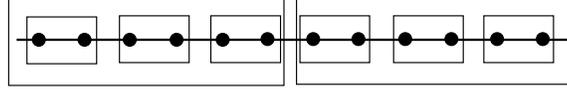}
\]
\caption{Definition of big and small cluster in $d=1$. $L=2$, $L_{c}=3$.
}
\label{emboitement.eps}
\end{figure}
We will then denote by $1 \leq \alpha, \beta \leq L$ the points in the
small cluster, by $1 \leq  A,B \leq  L_{c}$ the positions of the small
clusters in the big clusters, and by $\bar A, \bar B$ the positions of
the big clusters on the full lattice {\sl expressed in units of the original lattice}.
Any point on the big cluster can be described with a couple $(\alpha
,A)$ and any point on the lattice by a triplet $(\alpha ,A,\bar A)$.
Similarly, we denote by $-\pi \leq  k \leq  \pi$ an element of the
original lattice's Brillouin zone, by $-\pi/L \leq  K \leq  \pi/L$
an element of the  Brillouin Zone of the superlattice of the small
clusters, and by  $-\pi/  (LL_{c}) \leq  \bar K \leq  \pi/ (LL_{c})$
an element of the  Brillouin Zone of the superlattice of the big
clusters.
Finally, the reciprocal lattice of the small cluster are 
$K_{c}= \frac{2\pi n}{L},  n=0,\dotsc,L-1$, the same for
the big cluster considered as a cluster of $L_{c}$ small clusters, in
units of the original lattice,  are $\bar K_{c}= \frac{1}{L}\frac{2\pi p}{L_{c}}, p=0,\dotsc,L_{c}-1$.
Hence the reciprocal lattice points of the big cluster considered as a
cluster of $LL_{c}$ points are $K_{c} + \bar K_{c}$.
We solve the \DCA\  in real space with the big cluster of size $LL_{c}$ :
\begin{align}\label{app.DCA.1}
G_{\alpha A,\beta B} (\omega ) &= \sum_{\bar K} \dfrac{1}{\omega -
t^{cyc}_{(\alpha, A),(\beta, B)} (\bar K) - \Sigma_{\alpha A,\beta B} (\omega )}
\\
t^{cyc}_{(\alpha ,A),(\beta ,B)} ( \bar  K) &= \frac{1}{( LL_{c})^{d}}
\sum_{K_{c},\bar K_{c}} e^{i 
(K_{c} + \bar  K_{c} )\cdot  (x_{\alpha} + x_{A} - x_{\beta} - x_{B})}
t( \bar K + K_{c} + \bar K_{c})
\\
&= \frac{1}{L_{c}^{d}}
\sum_{\bar K_{c}} 
e^{i \bar  K_{c}\cdot (x_{\alpha} + x_{A} - x_{\beta} - x_{B})}
t^{cyc}_{\alpha \beta} (\bar K + \bar K_{c})
\end{align}
where we used $K_{c}\cdot x_{A} = 0 \ [2\pi ]$ in the last equation.
Because of the reduced invariance, we can diagonalise using $\bar
K_{c}$ and obtain : 
\[
G_{\alpha \beta} ( \bar K_{c},\omega ) = \sum_{\bar K} \dfrac{1}{\omega
-t^{cyc}_{\alpha ,\beta } (\bar K +\bar K_{c}) e^{i \bar K_{c} \cdot(x_{\alpha} -x_{\beta} )} -
\Sigma_{\alpha,\beta} (\bar K_{c},\omega )}
\]
Using the {\it unitary} transformation :
\[
\widetilde{G}_{\alpha \beta} (\bar K_{c},\omega) \equiv  e^{-i \bar
K_{c} \cdot(x_{\alpha} -x_{\beta} )}
 G_{\alpha \beta } (\bar K_{c}, \omega )
\]
we obtain :
\begin{equation}\label{app.DCA.2}
\widetilde{G}_{\alpha \beta} ( \bar K_{c},\omega ) = \sum_{\bar K} \dfrac{1}{\omega
-t^{cyc}_{\alpha ,\beta } (\bar K +\bar K_{c}) -
\widetilde{\Sigma}_{\alpha,\beta} (\bar K_{c},\omega )}
\end{equation}

Let us now concentrate on the AF order : $L=2$ and for simplicity in
dimension 1. In this case, $K_{c} = 0, \pi$. We will now show that
(\ref{app.DCA.2}) is equivalent to the $k$-space formulation  presented in
\cite{MaierPhD} (denoted by a upperscript $M$). In the AF phase, correlations appears
between $k$ and $k+ \pi $ and the self-consistency condition reads :
\begin{equation}\label{App.DCA.self.cond.Maier}
G^{M}( \bar K_{c},\omega ) = \sum_{\bar K} 
\parent{
\begin{pmatrix}
\omega - t (\bar K +\bar K_{c}) & 0 \\
0 & \omega - t (\bar K +\bar K_{c} + \pi )
\end{pmatrix}
- \Sigma^{M} (\bar K_{c},\omega )}^{-1}
\end{equation}
where $G^{M}$ and $\Sigma^{M}$ are $2\times 2$ matrices, which are non
diagonal in the AF phase : 
\[
G^{M} \equiv -
\begin{pmatrix}
\moy{T c_{k}c^{\dagger}_{k}} & \moy{T c_{k}c^{\dagger}_{k+ \pi }} \\
\moy{T c_{k+ \pi }c^{\dagger}_{k}} & \moy{T c_{k+\pi }c^{\dagger}_{k+ \pi }}
\end{pmatrix}
\]
Using the transformation \cite{MaierPhD}
\[
c_{1K} = \dfrac{c_{K}+c_{K+\pi }}{\sqrt{2}} \qquad
c_{2K} = \dfrac{c_{K}-c_{K+\pi }}{\sqrt{2}}
\]
Eq. (\ref{App.DCA.self.cond.Maier}) is equivalent to
(\ref{app.DCA.2}). Therefore the two formulations of DCA are equivalent
in the antiferromagnetic phase.

The real space formulation of \DCA\  share with \CDMFT\  a property
which is useful in practice : 
one doesn't have to {\it anticipate} the appearance of ordered phase, 
i.e. to adapt the cluster scheme for the order to be described.
The order will show up automatically solving 
the real space \DCA\  equations.
Moreover, it can be more complex than an AF order, as long as the
cluster is big enough to contain at least one unit cell.

\section{Large-$U$ limit of the effective action : proof of (\ref{Classical.cluster.scheme})}
\label{App.details.class.Ugrand}
Here, we present some details of the derivation of the large-$U$ limit
of Eq. (\ref{class.effectiveaction}), in the limit where 
$U\rightarrow \infty$ with  $\beta \rightarrow
\infty$, $\beta /U$ fixed. The  effective action is 
\begin{align}\label{App.class.effectiveaction}
  S_{\text{eff}} & = - \iint_{0}^{\beta } d \tau  d \tau'
c^{\dagger}_{\mu  \uparrow} (\tau )
G_{0, \mu \nu \uparrow }^{-1} (\tau ,\tau')
c^{ }_{\nu  \uparrow} (\tau' )
+
\int_{0}^{\beta } d \tau
c^{\dagger}_{\mu  \downarrow} (\tau )
\partial_{\tau} 
c^{ }_{\mu  \downarrow} (\tau)
+ 
U
\Bigl (n_{\mu \uparrow}(\tau )  - \frac{1}{2}\Bigr ) 
\Bigl (n_{\mu \downarrow}(\tau )- \frac{1}{2}\Bigr )
\\
G_{0\mu \nu\uparrow}^{-1}(i\omega_{n})&= i \frac{xU}{2} \delta_{\mu \nu }-
\tilde{t}_{\mu \nu } (x)
- \frac{2t}{U} J^{\mu \nu }_{\rho } (x)\moy{h_{\rho}}  + O \parent{\frac{1}{U^{2}}}
\end{align}
We are going to compute the expression of partial partition function $Z[\{n_{\rho \downarrow }
\}]$ (also denoted as $Z[\{S_{\rho } \}]$) as a function of the spins
and and recover (\ref{Classical.cluster.scheme}).
We first compute the expansion of $\ln Z[\{n_{\rho \downarrow } \}]$ at second order in $t$ and first
order in $J$. We obtain :
\begin{multline}\label{app.class.effect.act.exp1}
\ln \biggl ( Z[\{S_{\rho } \}]/Z_{0}[\{S_{\rho } \}] \biggr )=
 \sum_{(i,j)}\sum_{i\omega_{n}, i\omega'_{n}} \tilde{t}_{ij} (i\omega_{n}) 
\tilde{t}_{ji} (i\omega'_{n}) \moy{c^{\dagger}_{i\uparrow}
(i\omega_{n})c_{j\uparrow} (i\omega_{n})c^{\dagger}_{j\uparrow}
(i\omega'_{n})c_{i\uparrow} (i\omega'_{n})}_{S}
+ \\
2 \sum_{i}\sum_{i\omega_{n}} J^{ii}_{\rho } (x) \moy{h_{\rho}} \moy{c^{\dagger}_{i\uparrow}
(i\omega_{n})c_{i\uparrow}(i\omega_{n})}_{S}
 + O (t^{3},J^{2})
\end{multline}
where $(i,j)$ denotes the sum over couples, $\moy{\quad}_{S}$ is an
average of the Gaussian action at fixed $S$.
We see that 
\begin{equation}
Z_{0}[\{S_{\rho } \}] \equiv  Z_{t=J=0}[\{S_{\rho }
\}] \propto \prod_{\rho } \cosh \biggl (  \frac{\beta U}{4} S_{\rho }\biggr )
\end{equation}
is independent of $S_{\rho}$  (since $S_{\rho} = \pm 1$) and therefore
it will  be dropped in the following.
Using the relations : 
\GroupeEquations{
\begin{align}
\moy{c_{\rho }(i\omega_{n})c^{\dagger}_{\rho'}(i\omega'_{n})} &=
-G^{S}_{\rho \rho'} (i\omega_{n})\delta_{\omega_{n},\omega'_{n}}
\\
G^{S}_{\rho \rho'} (i\omega_{n}) & \equiv   \frac{\delta_{\rho \rho '}}{i
\omega_{n} + \frac{U}{2}S_{\rho }} =  
\frac{\delta_{\rho \rho '}}{2}
\left(
\frac{1-S_{\rho }}{i\omega_{n} - \frac{U}{2}} + \frac{1+S_{\rho }}{i\omega_{n} + \frac{U}{2}}
 \right)
\end{align}
}
and the fact that in the AF phase $\tilde{t}(-z)=\tilde{t} (z)$ since
$\moy{S_{i}} + \moy{S_{j}} =0$,
we can reduce the expression (\ref{app.class.effect.act.exp1}), extract the $S_{i} S_{j}$ term and
obtain finally :
\begin{equation}
\ln \bigl ( Z[\{S_{\rho } \}]\bigr ) = - \beta H_{\text{eff}} + O (t^{3}, J^{2})
\end{equation}
with $H_{\text{eff}}$ given by Eqs. (\ref{Classical.cluster.scheme}).
The additional term $J^{\mu \nu}_{\rho }$ does not contribute to the
$J_{\text{Ising}}$ since it would give a $O (1/U^{2})$ contribution.
Moreover, since $J_{\text{Ising}}\sim \tilde{t}^{2}/U$
we do not need to compute the $1/U$ term of $\sigmacluster$.

Finally, higher orders in the expansion do not contribute. Indeed,
since $Z[\{S_{\rho } \}]$ is a {\it gaussian} action  and since 
we compute $\ln Z[\{S_{\rho } \}]/Z_{0}$, we only get {\it connected
diagrams}, thus we have one sum over Matsubara frequencies. 
Each Matsubara sum gives a factor $\beta U$ (in the large-$U$ limit
$\sum_{\omega_{n}}$ is
replaced by $ \frac{\beta}{4\pi}\int d (xU)$) and  each $G$ a factor $U^{-1}$.  
Therefore higher order terms are subdominant.

\section{Classical limit of PCDMFT}\label{App.classical.calcul.pcdmft}
\def\B#1#2#3{\frac{\moy{h^{#1}_{#2}h^{#1}_{#3}}}{\moy{h^{#1}_{#2}}\moy{h^{#1}_{#3}}}}
\def\Bc#1#2#3{\frac{\moy{h^{#1}_{#2}h^{#1}_{#3}}_{c}}{\moy{h^{#1}_{#2}}\moy{h^{#1}_{#3}}}}
\def\BB#1#2#3{\frac{\moy{h^{#1}_{#2}h^{#1}_{#3}}_{c}}{\moy{h^{#1}_{#2}h^{#1}_{#3}}}}
\def\som{\frac{1}{S_{c}} \sum_{\empile{\rho \in {\cal C}}{\rho +\delta \in {\cal C}}}}
\def\somA{\frac{1}{S_{c}} \sum_{\empile{\rho \in {\cal C}_{A}}{\rho +\delta \in {\cal C}}}}
\def\somB{\frac{1}{S_{c}} \sum_{\empile{\rho \in {\cal C}_{B}}{\rho +\delta \in {\cal C}}}}
In this Appendix, we present the solution of  Eqs.
(\ref{Def.SigmaLatt.AF},\ref{Classical.Sigma.From.Seff},\ref{Classical.Expansion.Selfcons})
for completeness.
We restrict ourselves to nearest neighbor hopping.
We start by solving at order 1 in the $1/U$ expansion. From
(\ref{Classical.Sigma.From.Seff}) and
(\ref{Classical.Expansion.Selfcons}), we have for $\mu \neq \nu $
($\Delta_{\mu \mu} =0$):
\begin{equation}\label{app.class.pcdmft.eq1}
\Delta_{\mu \nu}^{\alpha } = {t}_{\nu -\mu } + 
\overline{\delta {\sigmalatt}_{\mu \nu }^{\alpha }} - \Delta_{\mu \nu }^{\alpha }
\Bc{\alpha }{\mu}{\nu}
\end{equation}
where $\alpha = 1,2$ is an index for the two (AF) solutions as in (\ref{Def.SigmaLatt.AF}).
Dividing the cluster into its two sublattices $A$ and $B$ with ${\cal
C} = {\cal  C}_{A} \cup {\cal C}_{B}$ and using
(\ref{Def.SigmaLatt.AF}), we have (for all $\mu \in {\cal C}$) :
\begin{equation}
\overline{\delta {\sigmalatt}^{\alpha }_{\mu, \mu +\delta }} = 
\somA \delta {\sigmacluster}^{\beta}_{\rho ,\rho +\delta} +
\somB \delta {\sigmacluster}^{\bar \beta}_{\rho ,\rho +\delta} 
\end{equation}
where $\beta =\alpha$ if $\mu \in {\cal C}_{A}$, $\beta =\bar \alpha$ otherwise,
since if $\rho +\delta \notin {\cal  C}$, the exponential depends on $K$  and
averages to zero.
For $\mu, \mu +\delta  \in {\cal  C}$, 
Eq. (\ref{app.class.pcdmft.eq1}) leads to :
\begin{equation}\label{app.class.pcdmft.eq4}
\Delta_{\mu ,\mu + \delta}^{\alpha } \B{\alpha }{\mu}{\mu +\delta}
 = t_{\delta} + 
\somA \Delta_{\rho ,\rho +\delta}^{\beta } \Bc{\beta }{\rho}{\rho
+\delta}
+
\somB \Delta_{\rho ,\rho +\delta}^{\bar \beta } \Bc{\bar \beta }{\rho}{\rho+\delta}
\end{equation}
for all $\mu ,\delta $. The left hand side depends only on whether
$\mu  \in {\cal  C}_{A}$ or not. Therefore, for all $\delta$, we can define for
$\mu \in {\cal  C}_{A}$ and $\zeta \in {\cal C}_{B}$ :
\GroupeEquations{
\begin{align}
F_{\delta}^{\alpha A } &\equiv \Delta_{\mu ,\mu + \delta}^{\alpha }
\B{\alpha }{\mu}{\mu +\delta}\\
F_{\delta}^{\alpha B } &\equiv \Delta_{\zeta  ,\zeta  + \delta}^{\alpha }
\B{\alpha }{\zeta }{\zeta  +\delta}
\end{align}}
(they are independent of $\mu$ and
$\zeta$). Eq. (\ref{app.class.pcdmft.eq4}) then leads to :
\GroupeEquations{
\begin{gather}
F_{\delta}^{\alpha A } = t_{\delta} +  A_{\delta \alpha } F_{\delta }^{\alpha A
}  + B_{\delta \alpha} F_{\delta }^{\bar \alpha B } \\
F_{\delta}^{\alpha B } = t_{\delta} +  A_{\delta \bar \alpha } F_{\delta
}^{\bar \alpha A}  + B_{\delta \bar \alpha} F_{\delta }^{ \alpha B } \\
 A_{\delta \alpha } \equiv  \somA \BB{\alpha }{\rho
}{\rho +\delta} \qquad 
B_{\delta \alpha } \equiv  \somB  \BB{\bar \alpha }{\rho }{\rho +\delta}
\end{gather}
}
 which can be rewritten as :
\GroupeEquations{
\begin{align}
\left(1 -  A_{\delta \alpha } \right)F_{\delta}^{\alpha A } -
B_{\delta \alpha}
F_{\delta }^{\bar \alpha B } &= t_{\delta} \\
\left(1 -  B_{\delta \alpha } \right)F_{\delta}^{\bar \alpha B } -
A_{\delta \alpha}
F_{\delta }^{ \alpha A } &= t_{\delta} 
\end{align}}
The determinant of the equations for $F$ is given by $\prod_{\alpha}
\bigl (  1 - A_{\delta \alpha } - B_{\delta \alpha }\bigr)$, which does not vanish
for a generic $x$. For $|\delta | >1$, $t_{\delta} = 0$ by
assumption hence $F_{\delta}^{\alpha A/B }=0$. 
In the classical limit, the two solutions are related by a spin-flip,
so $\BB{\alpha }{\mu }{\nu }$ does not depend on $\alpha$ for $\mu,\nu
$ nearest neighbor, according to (\ref{class.relationshS.AF}).
For $|\delta | =1$, 
the unique solution is obtained for equal $F_{\delta} \equiv F_{\delta}^{\alpha A/B }$,
which is given by :
\begin{equation}
F_{\delta} =
\dfrac{t_{\delta }}{\displaystyle  1-  \som \BB{}{\rho }{\rho +\delta} }
\end{equation}
Using the square symmetry, one can pick up one $\delta$ to compute the denominator.
Hence, we obtain Eq. (\ref{JIsingPCMDFT1}) and
(\ref{JIsingPCMDFT3}) (the hopping is nearest neighbors).
Note that the hopping is renormalised at order 1 (in $1/U$), but it is
still restricted to nearest neighbors.
In order to compute the $J$ term, we now focus on $ t + \delta
\sigmalatt$ which satisfies : 
\begin{equation}\label{app.class.pcdmft.eq2}
\overline{\bigl ( t + \delta \sigmalatt \bigr)}_{\mu, \mu  + \delta} = F_{\delta }
\end{equation}
It is a translation invariant quantity, which is also restricted to nearest neighbors.
It has a formula analogous to (\ref{App.DCA.t.supernotations}) for the
hopping, and its value inside the cluster is given by its $K$ average.
Using (\ref{Classical.Expansion.Selfcons.J}) to compute the mean
field  terms $J^{\mu \mu}_{\nu}$, we see that the computation is
similar than for \CDMFT, with just a renormalisation of the hopping
(note however that $t + \delta \sigmalatt$ is {\it not} the effective
hopping $\tilde{t}$ defined as the order 1 term of $\Delta$). In particular, the
$J_{B}$ terms have the same range as the original lattice hopping.
Reporting (\ref{app.class.pcdmft.eq2}) into (\ref{Classical.Expansion.Selfcons.J})
leads to Eq. (\ref{JIsingPCMDFT1}), which completes the derivation.

\section{Classical limit of the Pair scheme}\label{App.Classical.Pair.Scheme}

The first task is to generalise the pair scheme in the presence of
antiferromagnetic order.
Denoting the two sublattices by $A$ and $B$, we have the approximation
for $\Phi$  : 
\begin{equation}
\Phi_{pair} =(1-z)\left( \sum_{i\in A}\Phi_{1}[G_{ii}] + \sum_{j\in
B}\Phi_{1}[G_{jj}]\right)
\,+\,\sum_{\empile{<ij>}{ \empile{ i\in A}{j\in B }}  }\Phi_{2}[G_{ii},G_{jj},G_{ij}]
\end{equation}
We denote by $(1)$ and $(2)$ the one site and the two site model respectively.
We group the two one-site models into $2\times 2$ diagonal matrices,
and denote with an index $diag$ the diagonal part of a matrix.
To write the self-consistency condition, we take as unit cell one site
on $A$, one site on $B$. $K$ is now a vector in the corresponding reduced Brillouin
zone, presented on Figure \ref{BZ.eps}.
\begin{figure}[bt]
\[
       \figx{10cm}{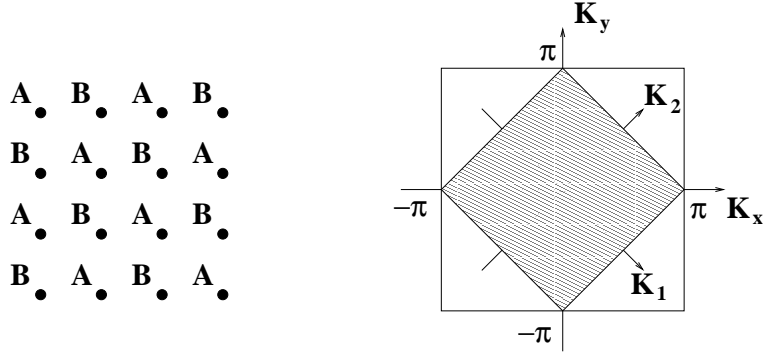}
\]
\caption{Notations and reduced Brillouin zone for the pair scheme
(shaded area)}
\label{BZ.eps}
\end{figure}
The hopping and the lattice self-energy are given by :
\GroupeEquations{
\begin{align}\label{app.ps.def.hopping.self}
t_{AA} (K) &= t_{BB} (K) = 0\\
t_{AB} (K) &= t C (K)\\
C (K) &= 1 + e^{-i (K_{1}- K_{2})} + e^{i (K_{1}-K_{2})} + e^{-i
(K_{1}+ K_{2})}\\
\sigmalatt_{diag} (K) &= \Sigma^{loc} = (1-z) \Sigma^{(1)}_{diag} + z
\Sigma^{(2)}_{diag} \\
\sigmalatt_{AB} (K) & =  \Sigma^{(2)}_{12}C (K)
\end{align}
}
and the Green function by the usual formula (with $2\times 2$
matrices).
\[
G = \overline{ (i \omega_{n} - t (K) - \sigmalatt (K,i \omega_{n}))^{-1}}
\]
The scheme implies a consistency equations for the Green functions : 
\begin{align*}
G^{(2)} &= G\\
G^{(1)} &= D (G)
\end{align*}
where we denote by $D$ the linear operator that restricts a matrix to
its diagonal.

First  $G_{0\downarrow}$ is diagonal (with same proof as for the other
schemes ) but here it is not trivial : 
\begin{align}
\bigl( G_{0\downarrow }^{(1 )} \bigr )^{-1} (i \omega_{n}) &=
i \omega_{n} +
\Sigma^{(1)}_{\downarrow }(i \omega_{n}) - \sigmalatt_{\downarrow }(i \omega_{n}) =i \omega_{n} -
z \left(\Sigma^{(2)}_{\downarrow }(i \omega_{n}) - \Sigma^{(1)}_{\downarrow }(i \omega_{n})
\right) \\
\bigl( G_{0\downarrow}^{(2 )} \bigr )^{-1} (i \omega_{n}) &=i \omega_{n} +
\Sigma^{(2)}_{\downarrow }(i \omega_{n}) - \sigmalatt_{\downarrow }(i \omega_{n}) =i \omega_{n} -
(z-1) \left(\Sigma^{(2)}_{\downarrow }(i \omega_{n}) - \Sigma^{(1)}_{\downarrow }(i \omega_{n})
\right) 
\end{align}
Using previous notations, we have :
\begin{equation}\label{app.ps.ratio.Delta.down}
\frac{\Delta_{\downarrow}^{(1)} (x)}{\Delta_{\downarrow}^{(2)} (x)} = \frac{z}{z-1}
\end{equation}

Let's turn now to the up electrons. Introducing the notations : 
\begin{align*}
\overline{A}^{\ (2)} &= \overline{A}\\
\overline{A}^{\ (1)} &= \overline{D (A)}
\end{align*}
and, using that, for $d$ a diagonal matrix independant of $K$,
$\overline{Ad}^{\ (\alpha) }= \overline{A}^{\ (\alpha )}d$ (for $\alpha
=1,2$),
we can expand the self-consistency conditions :
\begin{align}\label{app.ps.self.cons}
\left(G_{0}^{(1)} \right)^{-1} & = \left(G^{(1)} \right)^{-1} +
\Sigma^{(1)}\\
\left(G_{0}^{(2)} \right)^{-1} & = \left(G^{(2)} \right)^{-1} +
\Sigma^{(2)}
\end{align}
 with the expansion (\ref{Classical.Expansion.Selfcons}), to get : 
\GroupeEquations{\label{app.Classical.Expansion.Selfcons.ps}
\begin{align}
{\Delta}^{\ (\alpha)}_{ii} (x)&= \overline{t_{ii}}^{\ (\alpha)} + 
\Sigma_{ii}^{loc} - \Sigma^{(\alpha)}_{ii}
+
 \frac{2t}{U} J^{(\alpha )}_{ii \rho } (x)\moy{h_{\rho}^{(\alpha )}}  + O \parent{\frac{1}{U^{2}}}
\qquad \qquad (\alpha,i=1,2)\\
\label{app.ps.Classical.Expansion.Selfcons.J}
 J^{(\alpha )}_{\mu \nu\rho } (x) &\equiv  
\frac{1}{t}
\left(
\overline{ \tilde{t}_{\mu \rho } \tilde{t}_{\rho \nu }}^{\ (\alpha)} 
- 
\overline{\tilde{t}_{\mu \rho }}^{\ (\alpha)} 
\overline{\tilde{t}_{\rho  \nu }}^{\ (\alpha)} 
\right)\\
\tilde{t} (K,x) &\equiv t (K) + \sigmalatt (K,x) 
\end{align}}
Clearly, $\rho = \bar i$ with the notation $\bar 2 =1, \bar 1= 2$.
Moreover, 
\GroupeEquations{
\begin{align}
 \overline{\sigmalatt }_{diag}^{\ (1)}  - \Sigma^{(1)} &= z
 \left( \Sigma^{(2)}_{diag} -\Sigma^{(1)} \right)\\
 \overline{\sigmalatt }^{\ (2)}_{diag}  - \Sigma^{(2)} &= (z-1)
 \left( \Sigma^{(2)}_{diag} -\Sigma^{(1)} \right)
\end{align}}
and the $J$ terms are given by :
\GroupeEquations{
\begin{align}
J^{(1)} &= \frac{1}{t}
\sum_{R} 
\tilde{t}_{\mu \rho } (R)
\tilde{t}_{\rho \nu } (-R)
\\
J^{(2)} &= \frac{1}{t}
\sum_{R\neq 0} 
\tilde{t}_{\mu \rho } (R)
\tilde{t}_{\rho \nu } (-R)
\end{align}}
which implies
\GroupeEquations{
\begin{align}
J^{(1)} &= z J_{0}\\
J^{(2)} &= (z-1) J_{0}
\end{align}}
Therefore the equations for $\Delta$ simplify into : 
\GroupeEquations{\label{app.ps.deltaup}
\begin{align}
\Delta^{(1)}_{\uparrow ii} &= z  \left( \Sigma^{(2)}_{diag} -\Sigma^{(1)}
\right) + 2z J_{0} \moy{h_{\bar i}^{(1)}} \\
\Delta^{(2)}_{\uparrow ii} &= (z-1)  \left( \Sigma^{(2)}_{diag} -\Sigma^{(1)}
\right) + 2 (z-1) J_{0} \moy{h_{\bar i}^{(2)}} 
\end{align}}
Using that $m^{(1)} = m^{(2)}$, we have $\moy{h_{\bar i}^{(2)}}=
\moy{h_{\bar i}^{(1)}}$ at dominant order and therefore
\begin{equation}\label{app.ps.ratioDelta.up}
\frac{\Delta^{(2)}_{\uparrow ii}}{\Delta^{(1)}_{\uparrow ii}} = \frac{z-1}{z}
\end{equation}
From (\ref{app.ps.ratio.Delta.down},\ref{app.ps.ratioDelta.up}), using
the computation of the classical field from $\Delta$ presented in
Appendix \ref{App.details.class.Ugrand}, we obtain the {\sl classical
variation method} defined in (\ref{def.CVM},\ref{def.CVM2}).

Strictly speaking, we only prove here that, in the large-$U$ limit, if
there is a magnetic solution, it obeys the CVM equations. We have not
shown that such non zero solution exist. 
A priori, one could wish to push the semi-classical computation
further and {\it compute} the classical field explicitely to check
that it gives the values prescribed by the consistency of the
classical equations.
However, this is much harder to do  than in previous schemes for the
following reason : contrary to all other schemes studied in this paper, there is
no cancellation between the diagonal part of $\sigmalatt$ and of
$\sigmacluster$. Therefore, we have to compute correction to $\moy{h}$
to order $1/U^{2}$, in order to compute $\Delta$ to order $1/U$ (the
relation (\ref{class.relationshS}) is  a priori valid only at
dominant order).
Because $\beta \sim U$, we need $\Delta$ up to order $1/U^{3}$ to get
the such correction (like we need $\Delta$ at order $1/U$ to get a
classical field of order 1). All these difficulties are hidden in the
$\Sigma^{(2)}_{diag} -\Sigma^{(1)}$ term in (\ref{app.ps.deltaup}).

\section{Proof of the Cutkovsky-t'Hooft-Veltman equation}\label{AppendixKeldysh}

In this appendix, we prove the Cutkovsky-t'Hooft-Veltman formula using
the Keldysh method \cite{Keldysh}. We first briefly present our
conventions.

\subsection{Notations}\label{app.notations}
We denote with a $+$ the upper contour (going from $-\infty$ to
$\infty$) and a $-$ the lower contour.
The definition of the Keldysh propagators (for fermions) is:
\begin{eqnarray}\label{KP}
-i<T\phi (x,t)\phi ^{\dagger }(x',t')>&=&G^{++}(x,x',t-t')\\
-i<\tilde{T}\phi (x,t)\phi ^{\dagger }(x',t')>&=&G^{--}(x,x',t-t')\\
 i<\phi ^{\dagger } (x',t')\phi (x,t)>&=&G^{+-}(x,x',t-t')\\
-i<\phi (x,t)\phi ^{\dagger }(x',t')>&=&G^{-+}(x,x',t-t')
\end{eqnarray}
After Fourier transformation ($\hat f(\omega )=\int dt e^{i \omega
t}f(t)$), we have in particular the following useful relations : 
\GroupeEquations{
\begin{align}
\rho (x,x',\omega ) &= -\frac{1}{\pi} \Im G_{R} (x,x',\omega)\\
G^{+-}(x,x',\omega )&= 2i\pi  \rho (x,x',\omega )\theta (- \omega )
\label{app.defpm}\\
 G^{-+}(x,x',\omega )&= -2i\pi  \rho (x,x',\omega )\theta (\omega )
\label{app.defmp}\\
( G^{++}(x,x',\omega ))^{*}&=-G^{--}(x',x,\omega )
\end{align}}
To illustrate the diagrammatics, consider for example 
as an example a second order self-energy diagram in Fig. \ref{fig1.eps}.
\begin{figure}[hbt]
\[
       \figx{6cm}{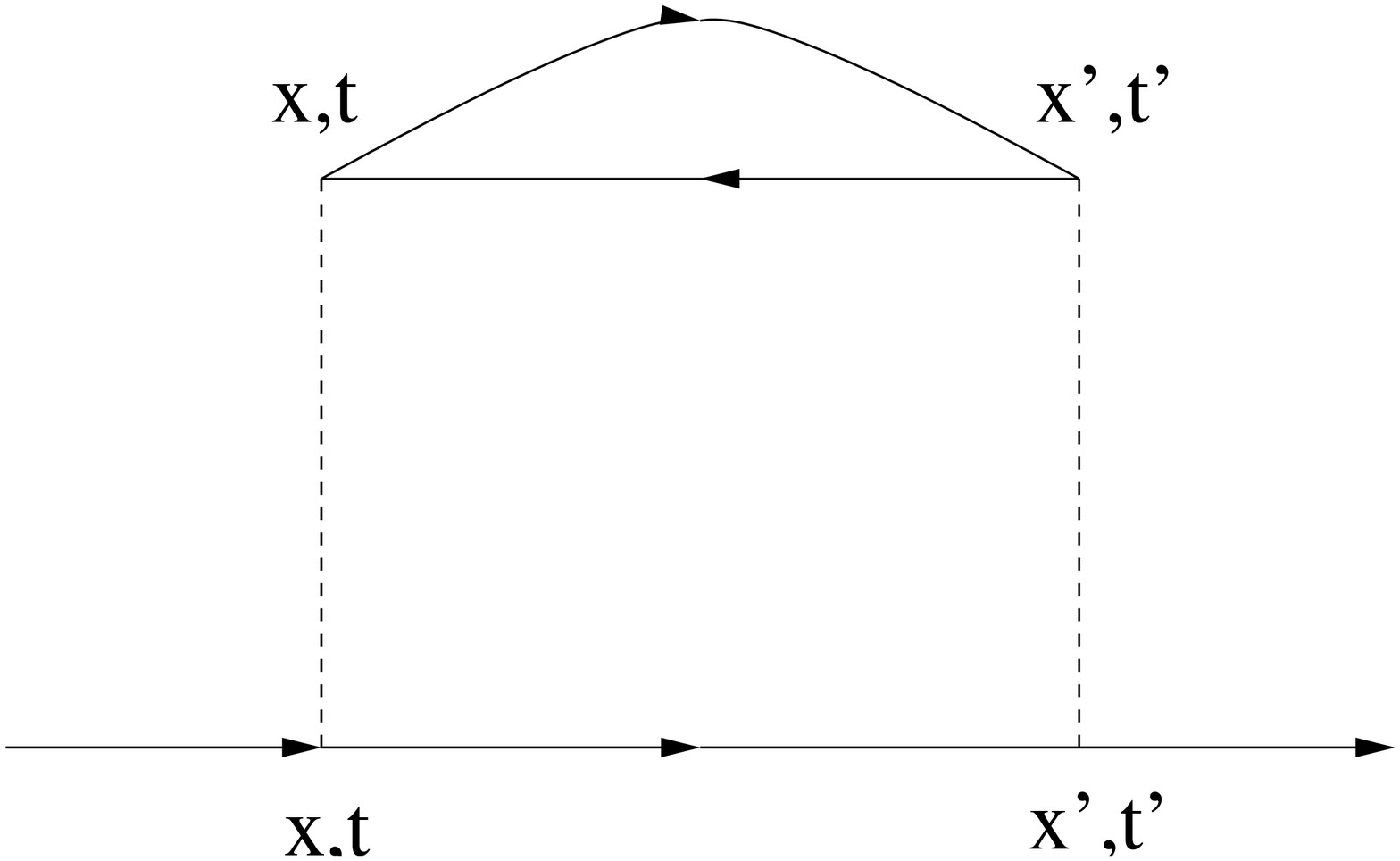} \hskip 1cm \figx{8cm}{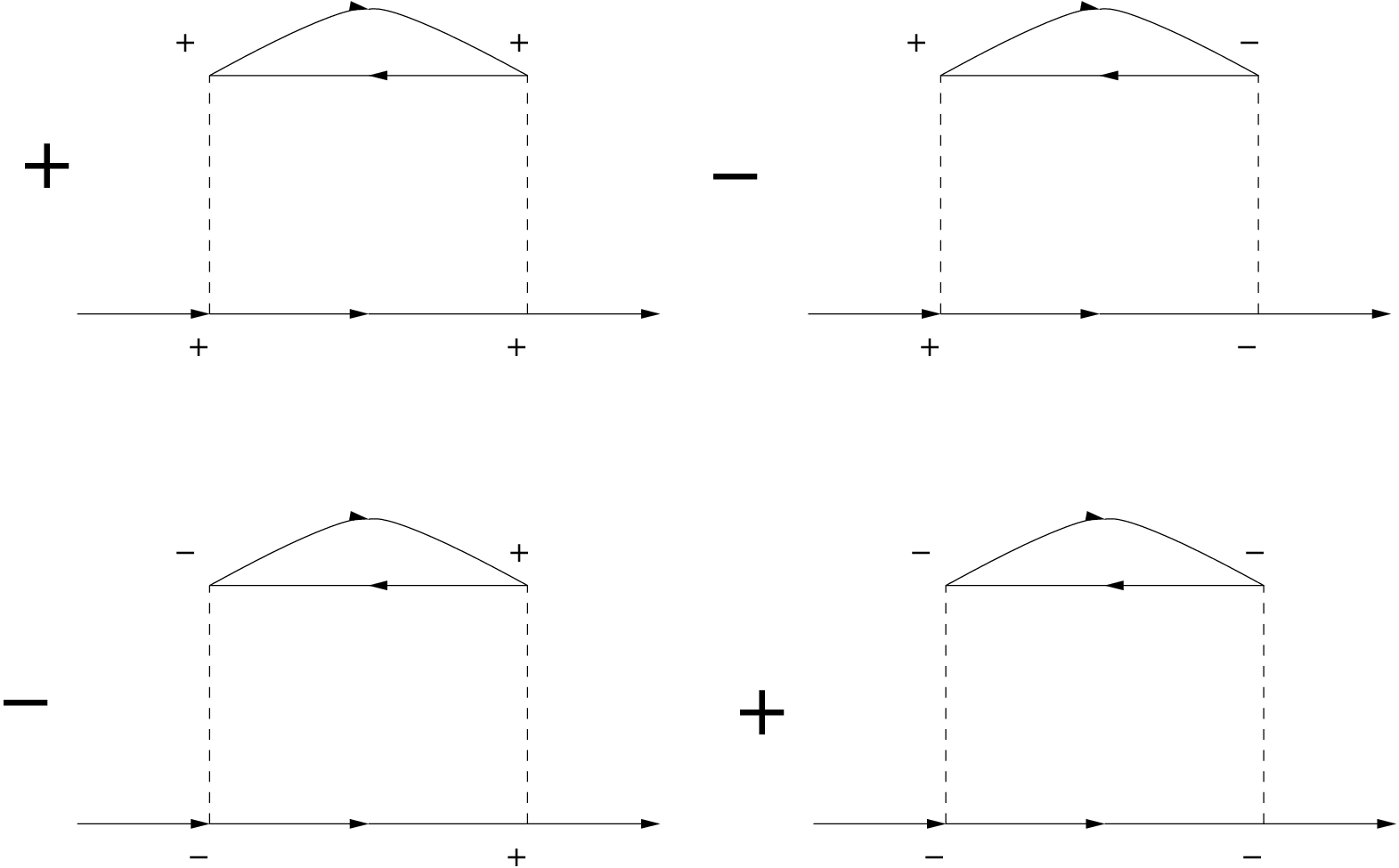}
\]
\caption{A self energy diagram and its Keldysh counterparts, with the
Keldysh indices and the overall factor.}
\label{fig1.eps}
\end{figure}
It corresponds to 
\[
G_{x,x',t-t'}G_{x,x',t-t'}G_{x',x,t'-t}
\]
The Keldysh counterparts of the diagram are 
obtained by replacing each vertex by his corresponding Keldysh
counterpart (See Fig \ref{fig1.eps}).
They corresponds to:
\[
G^{++}_{x,x',t-t'}G^{++}_{x,x',t-t'}G^{++}_{x',x,t'-t}-
G^{+-}_{x,x',t-t'}G^{+-}_{x,x',t-t'}G^{+-}_{x',x,t'-t}-
G^{-+}_{x,x',t-t'}G^{-+}_{x,x',t-t'}G^{-+}_{x',x,t'-t}+
G^{--}_{x,x',t-t'}G^{--}_{x,x',t-t'}G^{--}_{x',x,t'-t}
\]
where $G^{\pm , \pm }$ are the Keldysh propagators.

\subsection{Proof}
The first point is to write the imaginary part of the zero-temperature
retarded self energy in terms of the Keldysh components : 
\begin{equation}\label{app.ImsigR}
\Im \Sigma_{R} (\omega) = \frac{\Sigma_{+-} (\omega ) - \Sigma_{-+}
(\omega)}{2i}
\end{equation}
An important point is that this relation must hold {\sl for each
diagram individually}, in order to be compatible with any approximation
considered as a diagram summation.
Indeed at $T=0$ for $\omega >0$
\[
\Sigma_{R} (\omega) = \frac{1}{2i} (\Sigma_{++} (\omega ) + \Sigma_{+-} (\omega ))
\]
and using (\ref{app.defpm},\ref{app.defmp}), it is sufficient to prove
that 
\begin{equation}\label{app.crucial}
\Sigma^{++}+\Sigma^{--}+\Sigma^{+-}+\Sigma^{-+}=0
\end{equation}
is verified {\sl for each diagram individually}.
A way to prove that, which is actually in
Veltman's book\cite{VeltmanBook},
consists in noticing that each diagram with a $+$ at the largest time 
vertex cancels against the same diagram with a $-$ at this largest time vertex.
This happens because $G^{++}(t-t')=G^{-+}(t-t')$ and $G^{--}(t-t')=G^{+-}(t-t')$ for $t>t'$
so the only thing that changes between one diagram and the other is a minus sign 
(associated with the $-$ vertex and not to the $+$ vertex). 
This is called by Veltman {\it the largest time equation}. To unveil
this relationship we use the following notation: circle a vertex if it corresponds to 
$\phi ^{-}$ and do not circle it if it corresponds to $\phi ^{+}$. Then 
the relationship with Veltman rules to obtain the largest time equation will become clear.

To obtain the Keldysh expansion of $\Im \Sigma_{R}$, we can thus take
the expansion of $\Sigma$ (as defined with any perturbation theory)
and put Keldysh indices on it (as explained in paragraph
\ref{app.notations}). 
First, we will focus on $\omega >0$.
Then $\Sigma_{+-} (\omega)=0$, we only need to compute $\Sigma_{-+}$, {\it i.e.} the
diagram with a $+$ at the incoming vertex, and a $-$ at the outgoing
vertex.
The crucial properties of {\bf  zero-temperature} diagrams is the
following : due to the presence of $\theta$ functions 
in frequency for $G^{+-}$ and $G^{-+}$ (Cf. paragraph
\ref{app.notations})), the frequency  
flows always from the $+$ vertex to the $-$ vertex. 
This implies that a diagram is zero if it contains a $-$ vertex to which no external line is 
connected and that is surrounded by $+$ vertices because
of frequency conservation (the same is true if one replace $+$ with $-$). 
This cancellation generalizes to a connected set, or region, of $-$ 
to which no external line is connected and that is surrounded by $+$ vertices. 
%
%
Since we only have two external vertices, the non-vanishing diagrams
are those with a line that cut propagators dividing them into a $+$
region $L$ on the left (connected to the incoming vertex) and a $-$
region $R$ on the right (connected to the outgoing vertex). 
This line is the cut of the diagram and the
sum over Keldysh indices reduces to the sum over all possible cuts.
(see Fig. \ref{fig5.eps}).
In the $L$ (resp. $R$) region,  we have to use $G^{++}$
(resp. $G^{--}$). 
Since $G^{--}(x,x',\omega )= -(G^{++}(x',x,\omega ))^{*}$, the $R$
part gives $D_{MR}^{*}$ times $(-1)^{\mbox {\small number of internal
lines}}$, where $MR$ is the (left) diagram part obtained from $R$ by 
inverting all the arrows.
Finally, to each cut line going from right to left (left  to right) 
is associated a $G^{+-}$ ($G^{-+}$) propagator.
Hence  we obtain the third rule of the text : 
each cut lines going from left to right (right to left) is replaced
by $\rho (x,x',\omega )\theta(\omega )$ ($\rho (x,x',\omega
)\theta(-\omega )$).\\
So there are two last things that remains to be clarifed in order to get
the eq. (\ref{Cutkovsky.eq}): (1) that the symmetry factors are taking
into account correctly and (2) that the phase between the half left
and right diagram is the one leading to eq. (\ref{Cutkovsky.eq}).
Concerning the symmetry factors one has to be sure that
all the counting is done correctly. In order to do this
is useful to replace each half (or left) diagram with the sum of all half diagrams 
obtained from the original one permuting the cut lines in all the
possible ways. The symmetry factor that one has to attach to them is
exactly the one needed to get back the right symmetry factor for
the original diagram by gluing together the half-right and the
half-left diagrams. The ``gluing'' operations means attach to each 
$i$ line (see fig. \ref{fig6.fig}) of the half diagram the
corresponding $i$ line of the left diagrams. 
Indeed to get the symmetry factor of a given diagram
one has to write all the topologically equivalent way to obtain the
same diagram. This can be done starting from the left and from the
right, i.e. writing all the different ways to get the same
topologically equivalent left and right diagrams and then attach them
in all the possible ways that give rise to the original diagram.
Thus, the symmetry factor related to a half or left diagrams are just
all the different ways (contractions) that can be used to create
them.
In this way the operation of ``gluing'' together a half and left diagram in which all
the cut lines have been permuted produces $n!$ (n is the number of cut
lines) times the original uncut diagrams. The $1/n!$ in
eq. \ref{Causality.proof1} is there exactly to balance this redundant $n!$
term. 

Finally, let us focus on the phase between the right part of the diagram and
$D_{MR}^{*}$ is 
\begin{equation}
a= (-1)^{l_{R}} (-1)^{v_{\alpha R}} (-1)^{v_{\alpha R} (n_{\alpha }+1)} (-1)^{n_{loop}} (-i)^{n} (i)^{n-1}
\end{equation}
The first term comes from the sign in $G^{--}= -G^{++*}$, with
$l_{R}$ the number of internal lines inside the right part $R$; the
second comes from the $-$ Keldysh vertices, with $v_{\alpha R}$ the number of
vertices of type $\alpha$ in the right part; the third comes from the
$(-i)^{n_{\alpha }+1}$ factor of each vertices at $T=0$ with $n_{\alpha }$
the number of outgoing lines in the vertices of type $\alpha$ : this
factor change under conjugation; the fourth comes from the $n_{loop}$
broken loops in the cut diagrams (we restrict ourselves for fermions here);
the fifth and sixth terms comes from the difference between $G_{+-}$
and $\rho$ (in the cut propagators going from left to right and right
to left respectively).
The number of cut loops is given by $n_{loop}=n -1$.
Moreover, summing all lines ending at a vertex in the right part, we
have :
\begin{equation}
\sum_{\alpha} v_{\alpha R} n_{\alpha}  = l_{R} + n  \qquad [2]
\end{equation}
where $[2]$ is the reduction modulo 2.
Finally we get  $a=i$  which, combined with (\ref{app.ImsigR}) leads to the Eq. (\ref{Cutkovsky.eq}).
Similarly, for $\omega <0$, we have $\Im \Sigma_{R} =
\Sigma_{+-}/2i$. The left part has $-$ vertices, the right part $+$
vertices. Using a similar analysis, we get $a=-i$, which leads to Eq. (\ref{Cutkovsky.eq}).

\newcommand{\PRB}{Phys. Rev. B}\newcommand{\PRL}{Phys. Rev. Lett}\newcommand{\NPB}{Nucl. Phys.}\newcommand{\RMP}{Rev. Mod. Phys.}\newcommand{\ADV}{Adv. Phys.}

\end{document}